\documentclass[
	aps,prd,
	reprint,
	showkeys,
	amsmath,amssymb,
	nofootinbib,
	floatfix
	]{revtex4-2}
\usepackage[utf8]{inputenc}
\usepackage{aas_macros}
\pdfoutput=1
\usepackage{hyperref}
\hypersetup{ 
			colorlinks=true,
			pdfauthor={Georgios Valogiannis and Cora Dvorkin},
			pdftitle={WST TBD},
			citecolor=blue
			}
\usepackage{graphicx}
\usepackage{xcolor}
\usepackage{multirow}

\begin{document}

\title{Going Beyond the Galaxy Power Spectrum: \\an Analysis of BOSS Data with Wavelet Scattering Transforms}

\author{Georgios Valogiannis}
 \email{gvalogiannis@g.harvard.edu}
\author{Cora Dvorkin}
 \email{cdvorkin@g.harvard.edu}
\affiliation{
 Department of Physics, Harvard University, Cambridge, MA, 02138, USA\\
}

\begin{abstract}
We perform the first application of the wavelet scattering transform (WST) to actual galaxy observations, through a WST analysis of the BOSS DR12 CMASS dataset. We included the effects of redshift-space anisotropy, non-trivial survey geometry, systematic weights, and the Alcock-Paczynski distortion effect, following the commonly adopted steps for the power spectrum analysis. In order to capture the cosmological dependence of the WST, we use galaxy mocks obtained from the state-of-the-art {\tt ABACUSSUMMIT} simulations, tuned to match the anisotropic correlation function of the BOSS CMASS sample in the redshift range $0.46<z<0.60$. Using our model for the WST coefficients, as well as for the first 2 multipoles of the galaxy power spectrum, that we use as reference, we perform a likelihood analysis of the CMASS data. We obtain the posterior probability distributions of 4 cosmological parameters, $\{\omega_b,\omega_c,n_s,\sigma_8\}$, as well as the Hubble constant, derived from a fixed value of the angular size of the sound horizon at last scattering measured by the {\it Planck} satellite, all of which are marginalized over the 7 nuisance parameters of the Halo Occupation Distribution model. The WST is found to deliver a substantial improvement in the values of the predicted $1\sigma$ errors compared to the regular power spectrum, which are tighter by a factor of $3-5$ in the case of flat and uninformative priors and by a factor of $3-8$, when a Big Bang Nucleosynthesis prior is applied on the value of $\omega_b$. Our results are investigative and subject to certain approximations, which we discuss in the text. 
\end{abstract}

\maketitle

%%%%%%%%%%%%%%%%%%%%%%%%%%%%%%%%%%%%%%%%%%%%%%
\section{Introduction \label{sec:intro}}

The process of gravitational instability, modulated by the expansion of the universe and the set of fundamental interactions between its basic constituents, has led to the emergence of the large-scale structure (LSS) of the universe out of the primordial cosmic density field. The observed 3-dimensional (3D) distribution of matter is, as a result, a powerful probe of fundamental physics, which can reveal a wealth of information about the nature of the late-time accelerated expansion of the universe \citep{Copeland:2006wr}, the law of gravity at large scales \citep{Ishak:2018his,doi:10.1146/annurev-astro-091918-104423, Alam:2020jdv}, the nature of dark matter \citep{LSSTDarkMatterGroup:2019mwo}, the properties of massive neutrinos \citep{LESGOURGUES2006307,Dvorkin:2019jgs} and the physics of the early universe \citep{Chen_2016}. Aiming to tap into this valuable resource, a wide array of current and future cosmological surveys, such as the Dark Energy Spectroscopic Instrument (DESI) \citep{Levi:2013gra}, the {\it Vera C. Rubin} Observatory Legacy Survey of Space and Time (LSST) \citep{Abell:2009aa,Abate:2012za}, Euclid \citep{Laureijs:2011gra}, and the {\it Nancy Grace Roman} Space Telescope \citep{Spergel:2013tha}, among others, will trace the galaxies in the cosmic web with unprecedented levels of accuracy, potentially allowing us to explore and test the vast landscape of cosmological scenarios that propose to tackle these unresolved questions. 

On the theory front, the expected influx of data needs to be complemented by an associated theoretical framework to quantify and extract all possible cosmological information encoded in the LSS of the universe, a step that is underway. One of the most efficient ways to extract information out of an observed dataset is through the evaluation of the 2-point correlation function or its Fourier-space counterpart, the power spectrum. Combined with theoretical modeling and/or state-of-the-art simulations to capture the cosmological dependence of the target statistic and its covariance matrix, the values of the cosmological parameters that constitute a given scenario can be determined, up to a certain degree of accuracy. Despite serving as a useful first line of attack in problems of cosmological parameter inference, the power spectrum analysis is known to be incomplete, because it fails to capture a significant part of the information content in the LSS: the non-Gaussian part of the distribution sourced by the process of gravitational instability that drives structure formation \citep{PhysRevLett.108.071301}. In order to fully exploit the additional gains associated with tapping into the nonlinear regime of the LSS, one thus needs to evaluate higher-order statistics beyond the 2-point function. Despite significant theoretical progress made in this direction in the past decade \citep{10.1093/mnras/stv961,10.1093/mnras/stw2679,BERNARDEAU20021,Hahn_2020, Hahn_2021,PhysRevD.105.043517,Philcox:2021hbm}, including higher-order moments as part of a standard parameter inference scheme quickly becomes intractable, both from a computational standpoint, because of the sharply rising dimensionality of the resulting data vector, but also on the theoretical modeling front. In addition, in more challenging cases, such as that of a probability distribution with a heavy tail, even a complete description of all moments would fail to capture all available information, while at the same time amplifying outliers by raising the density field to very high powers \citep{PhysRevLett.108.071301}. 

Furthermore, realistic probes of the LSS, such as the 3D pattern of galaxies observed by spectroscopic surveys, do not generally perfectly trace the underlying matter distribution, but are rather \textit{biased} tracers of it \citep{DESJACQUES20181}, due to the complicated physics of structure formation. In addition, when galaxies are identified through spectroscopic means, their peculiar velocities around the Hubble flow lead to a perceived anisotropy of the observed clustering pattern, the redshift-space distortions (RSD) \citep{Hamilton1998}. Combined with other challenges associated with systematic errors or galaxies observed in a non-trivial survey geometry \citep{1994ApJ...426...23F}, an additional layer of complexity is added that makes the modeling and interpretation efforts more difficult. 

Aiming to overcome the former challenge, an active area of research is focused on developing novel estimators beyond the 2-point function, attempting to access higher-order information without having to explicitly evaluate the full correlation hierarchy. This broad spectrum of approaches consists of, but is not limited to, attempts to Gaussianize the density distribution by suppressing the contribution of LSS from regions with high overdensity \citep{Neyrinck_2009,PhysRevLett.108.071301,PhysRevLett.107.271301,White_2016,PhysRevD.97.023535,PhysRevLett.126.011301}, estimators that harness the information from regions that have not undergone nonlinear gravitational collapse, the cosmic voids \citep{Pisani:2019cvo,Massara_2015, 10.1093/mnras/stz1944, 10.1093/mnras/stv777, Hamaus_2015, Kreisch:2021xzq,Bonnaire:2021sie}, proxy lower-order statistics \citep{PhysRevD.91.043530,Dizgah_2020,Chakraborty:2022aok}, and a variety of other statistics beyond the power spectrum, such as Minkowski functionals \citep{10.1046/j.1365-8711.1999.02912.x,10.1111/j.1365-2966.2012.21103.x,10.1093/mnras/stt1316,PhysRevLett.118.181301}, the k-nearest neighbor cumulative distribution functions \citep{Banerjee:2020umh}, the minimum spanning tree \citep{Naidoo:2021dxz} or 1-point statistics \citep{Uhlemann:2019gni}. More recently, Convolutional Neural Networks (CNNs) \citep{6522407} have emerged as a completely novel approach that promises to reliably identify features of complex datasets, including a potentially full extraction of their non-Gaussian information content. Despite very promising results on cosmological applications \citep{PhysRevD.97.103515,Villaescusa_Navarro_2021, Perez:2022nlv,Dvorkin:2022pwo}, the extent to which their outcomes can be interpreted in order to allow reliable applications on real galaxy data is still an open question.

In our recent work \citep{Valogiannis:2021chp}, we investigated the prospect of bridging the gap between the use of traditional estimators and CNNs in modern cosmological analyses of the LSS, using the Wavelet Scattering Transform (WST) estimator \citep{https://doi.org/10.1002/cpa.21413,6522407}. Originally proposed in the context of signal processing in computer vision, the WST subjects an input physical field to a series of successive nonlinear operations (wavelet convolution and modulus), in a network architecture that resembles the one of a CNN with fixed kernels \citep{https://doi.org/10.1002/cpa.21413,6522407,Sifre_2013_CVPR,10.1214/14-AOS1276,6822556}. The outcome of this process is a compact set of a few coefficients that can serve as a basis that reflects the clustering properties of the input field beyond the 2-point function, without raising the field to very high powers, a common shortcoming of traditional estimators \citep{PhysRevLett.108.071301}, while at the same time retaining its interpretability, unlike in CNNs \citep{Sifre_2013_CVPR,10.1214/14-AOS1276,6822556}. In light of the great promise held in the use of this estimator, the WST has been recently applied in the fields of astrophysics \citep{refId0, Saydjari_2021,refdust}, cosmology \cite{10.1093/mnras/staa3165,Valogiannis:2021chp,Cheng:2021hdp,PhysRevD.102.103506,Greig:2022lfj,2022arXiv220407646E} and molecular chemistry \citep{10.5555/3295222.3295400,doi:10.1063/1.5023798} (a review can be found in Ref. \citep{cheng:2021xdw}).

Through the first WST application to 3D matter density fields, simulated by the {\tt QUIJOTE} simulations \citep{Villaescusa_Navarro_2020}, we showed in Ref. \citep{Valogiannis:2021chp} how the WST can deliver a very large improvement in the extracted errors on cosmological parameters. 
Motivated by these promising results, in this work we carry out the \textit{first} WST application to actual galaxy observations. In particular, we use the WST to analyze galaxy observations from the twelfth data release (DR12) \citep{Alam_2015} of the Baryon Oscillation Spectroscopic Survey (BOSS), a part of Sloan Digital Sky Survey, SDSS-III \citep{Eisenstein_2011,Dawson_2012}, and in particular the CMASS sample. We include the effects of redshift-space anisotropy, non-trivial survey geometry, the shortcomings of the dataset through a set of systematic weights and the Alcock-Paczynski effect, following the commonly adopted steps for the power spectrum analysis. In order to model the cosmological dependence of the WST coefficients extracted from galaxy observations, we make use of state-of-the-art simulated mocks that have been tuned to match the clustering properties of CMASS, through a set of free parameters absorbing the physics of galaxy formation, that we marginalize over. Using this framework, we perform a likelihood analysis of the BOSS CMASS dataset, which allows us to infer the values of cosmological parameters with the WST estimator. We finally compare our WST results against the ones obtained by the multipoles of the regular galaxy power spectrum, which we use as a reference, and clarify certain approximations in our analysis.

The rest of the paper is structured as follows: in \S\ref{WST:sup} we introduce the Wavelet Scattering Transform, and in \S\ref{sec:Analysis} we lay out all the details related to the analysis of the BOSS dataset, and of the associated mocks, using the WST and the power spectrum multipoles. We then present our results in \S\ref{sec:Results}, while concluding in \S\ref{sec:Conclusions}. More technical results are included in Appendices \S\ref{sec:App_WSTmask}, \S\ref{sec:App_convergence}, \S\ref{sec:App_compression} and \S\ref{sec:App}. 

%%%%%%%%%%%%%%%%%%%%%%%%%%%%%%%%%%%%%%%%%%%%%%
\section{Wavelet Scattering Transform}\label{WST:sup}

The Wavelet Scattering Transform estimator \citep{https://doi.org/10.1002/cpa.21413,6522407} was originally proposed in the context of signal processing in computer vision as a means of capturing the statistical properties of an input field. In addition to exhibiting a set of powerful and well-understood mathematical properties (superseding the ones of the conventional power spectrum) \citep{https://doi.org/10.1002/cpa.21413}, it was also shown to provide key insights into the nature of convolutional neural networks \citep{6522407}. As a result, and as we will see more clearly below, it can constitute an ideal middle-ground between these two types of approaches. 

In the WST approach, a given input field, $I(\bold{x})$, is subjected to two basic nonlinear operations: wavelet convolution and modulus. That is, if $\psi_{j_1, l_1}(\bold{x})$ is an oriented wavelet probing a scale $j_1$ and angle $l_1$, under the fundamental WST operation, $I(\bold{x})$ will be transformed as:
\begin{equation}\label{eq:convmod}
I'(\bold{x}) = |I(\bold{x}) \ast \psi_{j_1, l_1}(\bold{x}) |,   
\end{equation}
where $\ast$ denotes convolution. Taking the expectation value\footnote{In practice, this corresponds to taking the spatial average of the field.} of Eq. \eqref{eq:convmod} produces a WST coefficient, $S$, which is nothing else than a real number characterizing the field. Combined with a family of localized wavelets $\psi_{j_1, l_1}(\bold{x})$ probing a range of scales $j_1$ and angles $l_1$, successive applications of the above procedure give rise to a $\textit{scattering network}$, the WST coefficients, $S_n$, of which are given by the following relations up to order $n=2$:

\begin{align}\label{eq:WSTcoeff:base}
 S_0 &= \langle |I(\bold{x})|\rangle, \nonumber \\
 S_1(j_1,l_1) &= \left\langle |I(\bold{x}) \ast \psi_{j_1, l_1}(\bold{x}) | \right\rangle, \\
 S_2(j_2,l_2,j_1,l_1) &= \left\langle |\left(|I(\bold{x}) \ast \psi_{j_1, l_1}(\bold{x}) |\right) \ast \psi_{j_2, l_2}(\bold{x}) |\right\rangle \nonumber,
\end{align}
where in Eq. \eqref{eq:WSTcoeff:base} and hereafter the angular brackets, $\langle . \rangle$, denote averaging over the sample. 
Given that a convolution with an oriented wavelet reflects spatial and angular information about the input field, it can be intuitively understood how the WST coefficients of order $n$ encode clustering information analogous to the $2^n$-point correlation function \citep{https://doi.org/10.1002/cpa.21413,10.1214/14-AOS1276}. As such, the network in Eq. \eqref{eq:WSTcoeff:base} generates a compact basis of coefficients that can be used to partially characterize the higher-order clustering properties of a physical field, a task of particular interest to modern cosmology. Furthermore, the fundamental WST operations, namely convolution and modulus, together with its hierarchical architecture resemble a CNN with fixed kernels \citep{https://doi.org/10.1002/cpa.21413,6522407}. The combination of the above properties leads to a powerful estimator that can capture the non-Gaussian information content encoded in a physical field, similar to a CNN, but while retaining the desired interpretability of conventional clustering statistics (e.g., correlation function) through a basis of a few WST coefficients. As opposed to the regular clustering statistics, the WST can extract higher-order correlations without raising the target field to high powers \citep{cheng:2021xdw}, which is computationally more efficient. Lastly, the WST has demonstrated the ability to better access the information content carried in physical fields with heavy-tailed probability distributions \citep{cheng:2021xdw}, a case that is particularly challenging for higher-order moments to describe \citep{PhysRevLett.108.071301}. A pedagogical overview of various other properties of the WST (such as texture characterization or field generation) can be found in Ref. \citep{cheng:2021xdw}, whereas the formal mathematical description is discussed in detail in Refs. \citep{https://doi.org/10.1002/cpa.21413,6522407,Sifre_2013_CVPR,10.1214/14-AOS1276,6822556}.

The relations in Eq. \eqref{eq:WSTcoeff:base} can be generalized to allow for operations on a target field raised to a given power, $q$, in the following manner:
\begin{align} \label{eq:WSTcoeff:power}
 S_0 &= \langle |I(\bold{x})|^q\rangle, \nonumber \\
 S_1(j_1,l_1) &= \left\langle |I(\bold{x}) \ast \psi_{j_1, l_1}(\bold{x}) |^q \right\rangle, \\
 S_2(j_2,l_2,j_1,l_1) &= \left\langle |\left(|I(\bold{x}) \ast \psi_{j_1, l_1}(\bold{x}) |\right) \ast \psi_{j_2, l_2}(\bold{x}) |^q\right\rangle \nonumber,
\end{align}
where the choice of values of $q<1$ or $q>1$ emphasizes underdense or overdense regions, respectively, and $q=1$ recovers the basic WST case. In the first 3D WST application to the large-scale structure of the universe \citep{Valogiannis:2021chp}, it was shown that highlighting cosmic voids with values of $q<1$ led to a substantial increase in the information extracted on fundamental parameters, particularly the sum of the neutrino masses, matching and also exceeding the performance of the marked power spectrum \citep{PhysRevD.97.023535,Massara_2015}. In order to leverage this property, we choose to work with the relations in Eq. \eqref{eq:WSTcoeff:power}, a choice also adopted by the 3D molecular chemistry WST application of \citep{10.5555/3295222.3295400}.

The input WST field $I(\bold{x})$, that we will specify in the next section, can have an arbitrary number of dimensions, as far as the WST is concerned. In the context of the 3D LSS observations we will focus on in this work, $I(\bold{x})$ will be a 3D field. In general, a family of wavelets, $\psi_{j_1, l_1}(\bold{x})$, can be generated by performing dilations and rotations on a mother wavelet, which can similarly take various forms according to the desired application. In the 3D WST implementation of this analysis\footnote{This was first introduced in the context of molecular chemistry applications.}, the mother wavelet is a solid harmonic, multiplied by a Gaussian envelope, of the form
\begin{equation}\label{solid:sup}
\psi^{m}_{l}(\bold{x}) = \frac{1}{\left(2\pi\right)^{3/2}}e^{-|\bold{x}|^2 /2 \sigma^2}|\bold{x}|^l Y_l^m \left(\frac{\bold{x}}{|\bold{x}|}\right),
\end{equation}
where $Y_l^m$ are the familiar Laplacian spherical harmonics and $\sigma$ is the Gaussian width in units of the field pixels. The dilations are then described by the following re-scaling of the wavelet argument:
\begin{equation}\label{dil:sup}
\psi^{m}_{j_1, l_1}(\bold{x}) = 2^{-3 j_1}\psi^{m_1}_{l_1}(2^{-j_1} \bold{x}).
\end{equation}
If we sum over the index $m$, and consider $l$ to describe the angular information of the wavelet family, then the WST coefficients in this particular case will be given by:
\begin{align} \label{eq:WSTcoeff:sol}
 S_0 &= \langle |I(\bold{x})|^{q} \rangle, \nonumber \\
 S_1(j_1,l_1) &= \left\langle \left(\sum_{m=-l_1}^{m=l_1}|I(\bold{x}) \ast \psi^{m}_{j_1, l_1}(\bold{x}) |^{2}\right)^{\frac{q}{2}} \right\rangle, \\
 S_2(j_2,j_1,l_1) &= \left\langle \left(\sum_{m=1}^{m=l_1}|U_1(j_1,l_1)(\bold{x}) \ast \psi^{m}_{j_2, l_1}(\bold{x}) |^{2}\right)^{\frac{q}{2}} \right\rangle \nonumber,
\end{align}
with
\begin{equation}
 U_1(j_1,l_1)(\bold{x}) = \left(\sum_{m=-l_1}^{m=l_1}|I(\bold{x}) \ast \psi^{m}_{j_1, l_1}(\bold{x}) |^{2}\right)^{\frac{1}{2}}.
\end{equation}
 
 After performing the $1^{st}$ order convolution of the input field over a scale determined by $j_1$ in Eq. \eqref{eq:WSTcoeff:sol}, any information in scales smaller than that will be obscured in any subsequent convolutions of said field within the scattering network. Indeed, the 2D weak lensing (WL) application by Ref. \citep{10.1093/mnras/staa3165} found that $S_2$ coefficients with $j_2<j_1$ did not contribute any substantial cosmological information\footnote{We do note, nevertheless, that using equivariant wavelets on 2D fields the work of Ref. \citep{2021arXiv210411244S} did find some residual power to be carried in those usually discarded coefficients.}. As a result, we also choose to work with $S_2$ coefficients with only $j_2>j_1$, as we also did in \citep{Valogiannis:2021chp}.  We additionally point out that, as opposed to Eq. \eqref{eq:WSTcoeff:power}, the particular implementation \eqref{eq:WSTcoeff:sol} developed by Ref. \citep{10.5555/3295222.3295400}, that we follow, uses the same angular scale $l_2=l_1$ for the second order coefficient $S_2 \equiv S_2(j_2,j_1,l_1)$. Even though such a restriction most likely causes some loss of angular information, it significantly reduces the associated computational cost of the 3D WST evaluations, a tradeoff that was found to still perform very well in the 3D LSS application of \citep{Valogiannis:2021chp}. 
 
 To summarize, given an input 3D field with a resolution of {\rm NGRID} cells on a side, a number of total spatial dyadic scales $J$ (that can never exceed $\log_2({\rm NGRID})$) and total orientations $L$, the indices 
\begin{equation}
(j,l) \in([0,..,J-1,J],[0,..,L-1,L]),    
\end{equation}
give rise to a total of 
\begin{equation}\label{Stotal}
S_0+S_1+S_2=1+(L+1)(J^2+3J+2)/2    
\end{equation}
WST coefficients up to $2^{nd}$ order. The final choices that need to be determined for a WST evaluation are the values of the power $q$ and Gaussian width $\sigma$, that we will appropriately choose in the next section. 

We should note, at this point, that when working with isotropic input fields, dimensionality reduction techniques can further reduce the number of WST coefficients to work with. These include averaging over all $l$ orientations for a given spatial scale $j$, in order to construct isotropic coefficients \cite{10.1093/mnras/staa3165,refId0, Saydjari_2021} or less aggressive reduction techniques that aim to retain a larger degree of isotropy \cite{2021arXiv210411244S}. Given that in this work we will apply the WST on an \textit{anisotropic} physical field (as is the one determined by galaxy observations in redshift space), we will not consider this reduction. Lastly, we note that the WST coefficients are commonly normalized as follows:
\begin{align} \label{eq:WSTnorm:sup}
\bar{S}_0 &= \log(S_0), \nonumber\\
\bar{S}_1 &= \log(S_1/S_0), \\
\bar{S}_2 &= \log(S_2/S_1), \nonumber 
\end{align}
a choice adopted by several past applications \citep{10.1214/14-AOS1276,refId0, Saydjari_2021,10.1093/mnras/staa3165,Cheng:2021hdp}. Despite the fact that the WL WST applications on 2D shear maps \cite{10.1093/mnras/staa3165,Cheng:2021hdp} found this re-normalized basis to break degeneracies between $\Omega_m$ and $\sigma_8$, our previous 3D application to cosmological density fields of Ref. \citep{Valogiannis:2021chp} did not find any noticeable gains in information associated with this basis. As a consequence, we choose to work with the bare WST coefficients given by Eq. \eqref{eq:WSTcoeff:sol}.

%%%%%%%%%%%%%%%%%%%%%%%%%%%%%%%%%%%%%%%%%%%%%%
\section{Analysis}\label{sec:Analysis}

In this section, we lay out the details of the particular galaxy dataset and mocks in our analysis, as well as of the procedure we follow in order to extract the WST and power spectrum estimators out of them in each case.

%%%%%%%%%%%%%%%%%%%%%%%%%%%%%%%%%%%%%%%%%%%%%%
\subsection{Dataset}\label{sec:dataset}

This works uses galaxy observations from the twelfth data release (DR12) \citep{Alam_2015} of the BOSS\footnote{All BOSS data, as well as the accompanied covariance PATCHY mocks, are publicly available at \url{https://data.sdss.org/sas/dr12/boss/lss/}.}, a part of Sloan Digital Sky Survey, SDSS-III \citep{Eisenstein_2011,Dawson_2012}. Specifically, we work with CMASS data in the redshift range $0.46<z<0.60$ that were observed from two separate parts of the sky, the North (NGC) and South Galactic Cap (SGC). In light of the fact that these are two distinct subsets of observations, we hereafter evaluate our clustering statistics on each one of them separately and then obtain the average, weighted by the corresponding values of their angular footprint, following the standard procedure in past analyses of BOSS data \citep{10.1093/mnras/stw1096,10.1093/mnras/stw3298}. To be specific, if $X_{\rm NGC}$ and $X_{\rm SGC}$ is our statistic (be it WST or $P(k)$ multipoles) evaluated from the NGC and SGC parts, respectively, then the resulting data vector used in our analysis is always given by:
\begin{equation}\label{eq:NplusS}
X_{\rm N+S} = \frac{\left(A_{\rm NGC}X_{\rm NGC}+A_{\rm SGC}X_{\rm SGC}\right)}{(A_{\rm NGC}+A_{\rm SGC})},
\end{equation}
where $A_{\rm NGC}=6851 \ {\rm deg}^2$ and $A_{\rm SGC}=2525 \ {\rm deg}^2$. We should note, at this point, that even though the redshift range of the entire CMASS sample commonly adopted in BOSS analyses is actually $0.43<z<0.70$, we work with a narrower z cut because this was the one used for the production of our galaxy mocks, as we will see in the next section. Due to the same reason, we only work with the CMASS, rather than also with the LOWZ BOSS sample, even though our WST estimator can be straightforwardly applied to any set of galaxy observations.

Each dataset is accompanied by a corresponding random unclustered catalogue, with the exact same angular footprint and selection function, in order to enable the evaluation of clustering statistics (as we will see below). We choose to work with the random catalogue that has a number density $50 \times$ greater than the one of the corresponding observed samples, a choice commonly adopted in previous analyses \citep{10.1093/mnras/stw1096,10.1093/mnras/stw3298}.

We now proceed to explain the procedure followed to evaluate the fractional overdensity field from the BOSS dataset, which serves as the fundamental quantity of interest needed to extract both the WST and the P(k) statistic. To do so, we start by converting the observed galaxy sky coordinates, right ascension (RA), declination (DEC) and redshift $z$ into comoving Cartesian coordinates, $x,y,z$, always assuming a cosmology of $\Omega_m=0.3152,h=0.6736$,  which, as we will see below, will correspond to our chosen fiducial cosmology. We account for the potential errors introduced when assuming an incorrect cosmology to perform this conversion, which are known as the Alcock–Paczynski (AP) distortion \citep{1979Natur.281..358A}. We will explain our strategy in \S\ref{sec:Pkeval} for the power spectrum and in \S\ref{sec:WSTeval} for the WST.  Using the publicly available package {\tt nbodykit}\footnote{\url{https://nbodykit.readthedocs.io/en/latest/index.html}}, we then proceed to embed the sample into a cubic box with a comoving side equal to $L=2820$ $\rm Mpc/h$, which corresponds to the smallest possible cube that can embrace our (irregularly shaped) sample. Finally, the desired quantity is the (weighted) fractional overdensity field of data in a realistic survey format, commonly referred to as the Feldman-Kaiser-Peacock (FKP) field, $F(r)$ \citep{1994ApJ...426...23F}, which we evaluate on a mesh through the following relationship:
\begin{equation}\label{fkpfield}
F(\bold{r})=\frac{w_{\rm FKP}(\bold{r})}{I_2^{1/2}}\left[w_c(\bold{r})n_g(\bold{r})-\alpha_r n_s(\bold{r}) \right].
\end{equation}
Here $n_g(\bold{r})$ and $n_s(\bold{r})$ are the observed number density of the galaxies and the objects of the random catalogue, respectively, and $\alpha_r$ denotes the ratio between the (weighted) total number of objects in the galaxy catalogue over the corresponding value of the synthetic random one. The BOSS dataset further includes 3  weights that reflect the incompleteness of the observed sample: a redshift failure weight, $w_{\rm rf}$, a fiber collision weight, $w_{\rm fc}$ and a systematics weight, $w_{\rm sys}$. They enter Eq. \eqref{fkpfield} as a combined contribution \citep{10.1093/mnras/stw1096,10.1093/mnras/stw3298}
\begin{equation}\label{eq:wc}
w_c(\bold{r}) = \left(w_{\rm rf}(\bold{r})+ w_{\rm fc}(\bold{r}) - 1.0 \right)w_{\rm sys}(\bold{r}).
\end{equation}
The remaining weight in Eq. \eqref{fkpfield} is the FKP weight \citep{1994ApJ...426...23F}, 
\begin{equation}\label{eq:fkpweight}
w_{\rm FKP}(\bold{r})=\left[1+\bar{n}_g(\bold{r})P_0\right]^{-1}, 
\end{equation}
for $P_0=10^{-4} \ {\rm Mpc}^3/h^3$, which is meant to ensure optimal extraction of information at small scales and is also provided for each galaxy of the sample. Finally, the normalization factor 
\begin{equation}\label{eq:Inorm}
I_2 = \int d^3\bold{r} \ w_{\rm FKP}^2 (\bold{r}) \langle w_c(\bold{r})n_g(\bold{r}) \rangle ^2
\end{equation}
is meant to normalize the amplitude of the power spectrum with respect to the observed power in an instance of no survey selection. It is straightforward to see that in the absence of a weighting scheme, Eq. \eqref{fkpfield} gives the regular galaxy overdensity field evaluated from a sample. The FKP field from Eq. \eqref{fkpfield} is the fundamental quantity of interest to extract from the data, which, as we will see below, can either be fed into Eq. \eqref{eq:WSTcoeff:sol} to obtain the observed WST coefficients or get Fourier transformed in order to obtain the multipoles of the anisotropic galaxy power spectrum. 

It should be noted, at this point, that the weighting scheme \eqref{eq:wc} was designed to account for the impact of the dataset incompleteness on the power spectrum, rather than the WST. As a result, it is possible that a different set of weights is needed in order to fully account for these effects on the WST coefficients. Given however, that the WST (partly) contains clustering information comparable to that in the 2-point correlation function, we expect Eq. \eqref{eq:wc} to at least partially capture the shortcomings of the dataset, from a WST point of view, and consider it a reasonable starting point for this first WST application. We defer a more detailed investigation of how to optimally model this effect for the WST to future work. Likewise, even though the FKP weights from Eq. \eqref{eq:fkpweight} were designed to ensure the optimal recovery of information by the power spectrum, we include them in the WST analysis as well, in order to maintain consistent inputs across our pipeline.

%%%%%%%%%%%%%%%%%%%%%%%%%%%%%%%%%%%%%%%%%%%%%%
\subsection{Mocks}\label{sec:mocks}

Previous analyses of BOSS data use more traditional estimators, which mostly relied on perturbation theory models of various kinds to capture the cosmological dependence of the target clustering statistics (with a few representative examples being \citep{10.1093/mnras/stw1096,10.1093/mnras/stw3298,Ivanov_2020,d_Amico_2020,Philcox_2020,PhysRevD.105.043517}). On the other hand, given the lack of a first principles theory model in place to predict its cosmological dependence, the WST approach demands the use of a full set of simulated mocks. In this section we introduce the set of mocks we will use for our model and covariance matrix for the WST (and the power spectrum, for comparison) in our final likelihood analysis.

%%%%%%%%%%%%%%%%%%%%%%%%%%%%%%%%%%%%%%%%%%%%%%
\subsubsection{\tt ABACUSSUMMIT mocks}\label{sec:Mocksabac}

In order to model the cosmological dependence of our estimators, we use the publicly available suite of {\tt ABACUSSUMMIT} simulations \citep{10.1093/mnras/stab2484}\footnote{Detailed information on all the simulations, as well as on how to access them, can be found at \url{https://abacussummit.readthedocs.io/en/latest/index.html}}, which were performed using the state-of-the-art {\tt ABACUS} N-body code \citep{10.1093/mnras/stz634,10.1093/mnras/stab2482}. Having evolved $6912^3$ dark matter particles in a cubic box of a side equal to $L_{\rm Box}=2000$  Mpc/h (for the `base' configuration), corresponding to a particle mass resolution $M_p = 2.1\times 10^9$ $M_{\odot}/h$\footnote{$h$ is the dimensionless Hubble constant, $h=H_0/(100$ km s$^{-1}$Mpc$^{-1}$).}, these simulations are able to not only match but also exceed the requirements of DESI \citep{Levi:2013gra}, making them the best available set of simulations to work with. The gravitationally bound dark matter halos in the {\tt ABACUSSUMMIT} collection were identified using the state-of-the-art spherical overdensity halo finder, {\tt COMPASO} \citep{10.1093/mnras/stab2980}.

We will now briefly summarize the cosmological landscape of the {\tt ABACUSSUMMIT}, starting with the fiducial `base' configuration, which corresponds to the mean values of the `base\_plikHM\_TTT\_EEE\_lowl\_lowE\_lensing' version of the {\it Planck} 2018 \citep{Planck2018} $\Lambda$CDM cosmology. In order to enable averaging over the effects of cosmic variance, the base cosmology was also run using $24$ additional different random initial phases, in addition to the base box. We begin by considering constraints around 4 cosmological parameters, $\{\omega_b,\omega_c,n_s,\sigma_8\}=\{0.02237,0.120,0.9649,0.8114\}$, where $\omega_X=\Omega_X h^2$ and $\sigma_8$ refers to the amplitude of density fluctuations traced by the combination of cold dark matter and baryons, `cb', in the presence of massive neutrinos with $\omega_{\nu}=0.0006442$. To capture the cosmological dependence on the above 4 parameters, we further consider additional simulations which vary each one of the parameters, in turn, and in a step-wise fashion, away from the fiducial background while keeping the rest fixed. This enables the evaluation of first-order derivatives. The exact list (as well as the associated parameter variations of these `first-order derivative grid' cosmologies) are listed in Table \ref{table:1}, while we also add that these were phase-matched to the base box, in order to cancel out the effects of cosmic variance upon taking central differences. All simulations have kept the value of the angular size of the sound horizon at last scattering, $\theta_{\star}$, fixed to the corresponding value derived from measurements by the {\it Planck} satellite \citep{Planck2018}: $100 \theta_{\star}=1.041533$, which implies a corresponding value of $h=0.6736$, for the base cosmology. Even though the Hubble constant was not explicitly varied in {\tt ABACUSSUMMIT}, the choice of a fixed $\theta_{\star}$ enables the evaluation of $h$ as an additional derived parameter, an option that we will consider below. The values of all the other cosmological parameters, which we keep fixed, together with a detailed description of the {\tt ABACUSSUMMIT} simulations can be found in Ref. \citep{10.1093/mnras/stab2484}.

\begin{table}[h!]
\centering
\begin{tabular}{|p{1.3cm}|p{1.3cm}|p{1.3cm}|p{1.3cm}|}
\hline 
$\omega_b$&$\omega_c$&$n_s$&$\sigma_8$\\
\hline \hline
 $0.02237$&$0.1200$&$0.9649$&$0.8114$\\
\hline
$0.02282$&$0.1200$&$0.9649$&$0.8114$\\
\hline
$0.02193$&$0.1200$&$0.9649$&$0.8114$\\
\hline
$0.02237$&$0.1240$&$0.9649$&$0.8114$\\
\hline
$0.02237$&$0.1161$&$0.9649$&$0.8114$\\
\hline
$0.02237$&$0.1200$&$1.0249$&$0.8114$\\
\hline
$0.02237$&$0.1200$&$0.9049$&$0.8114$\\
\hline
$0.02237$&$0.1200$&$0.9649$&$0.8698$\\
\hline
$0.02237$&$0.1200$&$0.9649$&$0.7532$\\
\hline
\end{tabular}
\caption{A list of all 8 {\tt ABACUSSUMMIT} first-order derivative cosmologies we use in this work, together with the corresponding values of the 4 cosmological parameters varied. The first row corresponds to the base cosmology, shown for reference.}
\label{table:1}
\end{table}

In order to generate realistic galaxy mock samples from the underlying dark matter and halo catalogues, {\tt ABACUSSUMMIT} uses the Halo Occupation Distribution (HOD) framework through a flexible package called {\tt ABACUSHOD} \citep{10.1093/mnras/stab3355}\footnote{The package is publicly available as part of {\tt ABACUSUTILS} \url{http://https://github.com/abacusorg/abacusutils}, together with an associated instruction manual at \url{https://abacusutils.readthedocs.io/en/latest/hod.html}. In particular, we use the {\tt ABACUSUTILS} version 1.0.4.}. In the baseline HOD implementation \citep{Zheng_2007}, simulated halos host galaxies based on a semi-analytical probablistic model that depends on 5 parameters. In particular, if $M_{\rm cut}$ denotes the minimum mass of a halo that can host a central galaxy, $\kappa M_{\rm cut}$ the minimum halo mass to host a satellite galaxy and $M_1$ characterizes the typical halo mass that hosts one satellite galaxy, the mean expected number of central, $\bar{n}_{\rm cent}(M)$, and satellite, $\bar{n}_{\rm sat}(M)$, galaxies assigned to a halo of mass $M$ are given by:
\begin{equation}\label{eq:ncent}
\bar{n}_{\rm cent}(M) = 0.5 \ {\rm erfc} \left[\frac{\log_{10}(\frac{M_{\rm cut}}{M})}{\sqrt{2} \sigma}\right],
\end{equation}
and
\begin{equation}\label{eq:nsat}
\bar{n}_{\rm sat}(M) = \left[\frac{M-\kappa M_{\rm cut}}{M_1}\right]^{\alpha} \bar{n}_{\rm cent}(M).
\end{equation}
The parameters $\alpha$ and $\sigma$ calibrate the relations \eqref{eq:ncent} and \eqref{eq:nsat}, and fully characterize the standard HOD model. We note that Eqs. \eqref{eq:ncent} and \eqref{eq:nsat} are applicable to luminous red galaxies (LRGs) \cite{10.1093/mnras/stv2382,10.1093/mnras/stw1014}, which mostly dominate the CMASS sample that we will work with. Out of a rich variety of extensions going beyond the vanilla HOD framework described above, we will adopt a version that accounts for the effect of the velocity bias of LRGs, a step shown to be necessary for an accurate fit against both BOSS data \citep{10.1093/mnras/stab3355} and hydrodynamical simulations \citep{10.1093/mnras/stac830}. The two additional HOD parameters it introduces are the central velocity bias, $\alpha_c$, which is meant to account for the mismatch between the velocity of central galaxies and the one of halo centers, and the satellite velocity bias, $\alpha_s$, which captures the equivalent effect for satellite galaxies. For a more in-depth discussion on more sophisticated HOD parametrizations (such as, for example, including assembly bias \citep{10.1093/mnras/stab235}) interested readers are referred to the relevant works referenced above. 

We can finally proceed to use our adopted 7-parameter HOD framework in order to generate galaxy mocks for our BOSS analysis. In particular, we closely follow the procedure laid out in Refs. \citep{10.1093/mnras/stab3355,Yuan:2022jqf} and search for an {\tt ABACUS}-derived galaxy mock that best fits the redshift-space 2-point correlation function, $\xi(r_{\perp},r_{\parallel})$, of the BOSS DR12 CMASS sample introduced in \S\ref{sec:dataset},  (averaged over North and South according to Eq. \eqref{eq:NplusS}), for the base cosmology. Here $r_{\perp},r_{\parallel}$ are the separations perpendicular and parallel to the line of sight, respectively, and we fit the CMASS correlation function using 8 logarithmic bins in the range 0.169-0.30 Mpc/h for $r_{\perp}$ and using 6 linearly spaced bins between 0 and 30 Mpc/h for $r_{\parallel}$, for the redshift cut $0.46<z<0.60$. The HOD parameters giving the best fit through this procedure, that we take as the fiducial HOD parameters, correspond to the following values:
\begin{align} \label{eq:HODfid}
&\{\alpha,\alpha_{\rm c},\alpha_{\rm s},\kappa,\log M_1,\log M_{\rm cut},\sigma\} = \\
& \{0.9022,0.2499,1.1807,0.3288,14.313,12.8881,0.02084\} \nonumber,
\end{align}
where $M_{\rm cut}$ and $M_1$ are expressed in units of $M_{\odot}/h$. We then proceed to use {\tt ABACUSHOD}, always with the parameters in Eq. \eqref{eq:HODfid} as input, so as to produce CMASS mocks for each one of the 8 first-order derivative cosmologies in Table \ref{table:1}, as well as for the 25 boxes of the fiducial cosmology. In order to null out any residual effects of cosmic variance during the evaluation of the derivatives, in addition, we generate 20 different HOD realizations for each one of the 8 derivative cosmologies. Finally, in order to capture the effects of varying the values of the HOD parameters, and in a direct analogy to the procedure followed for the 4 cosmological parameters, we proceed to generate first-order derivative mocks for HOD variations. We kept all parameters fixed to the fiducial values and successively varied one HOD parameter at a time, in a step-wise fashion. With $\theta_{\rm HOD}$ being the vector of HOD parameter values from Eq. \eqref{eq:HODfid}, we take steps $\theta_{\rm HOD} \pm \Delta \theta_{\rm HOD}$ as follows:
\begin{align} \label{eq:DelHODfid}
&\{\Delta \alpha,\Delta \alpha_{\rm c},\Delta \alpha_{\rm s},\Delta \kappa,\Delta \log M_1,\Delta \log M_{\rm cut},\Delta \sigma\} = \\
& \{0.36088, 0.02499, 0.11807, 0.2959, 0.7157,
       0.3866, 0.020\} \nonumber.
\end{align}
The resulting output of the procedure described in this section is a collection of 25 mocks for the base cosmology and a total of $20\times 2 \times 11 = 440$ mocks for the cosmological derivative variations, which constitute the full set of simulations that we will use in our analysis. 

%%%%%%%%%%%%%%%%%%%%%%%%%%%%%%%%%%%%%%%%%%%%%%
\subsubsection{{\tt PATCHY} mocks}\label{sec:Patchy}

In order to construct a Gaussian likelihood for our inference framework, that we will explain in detail in \S\ref{sec:mcmc}, we also need to construct an accurate covariance matrix in addition to the model for the observable. To do so, we use the $2048$ realizations of the publicly available {\tt MULTIDARK-PATCHY} mocks\footnote{Available at \url{https://data.sdss.org/sas/dr12/boss/lss/dr12_multidark_patchy_mocks/}.} \citep{10.1093/mnras/stv2826,10.1093/mnras/stw1014}, hereafter referred to as {\tt PATCHY} mocks, a collection large enough to make them ideal candidates for the evaluation of a properly converged covariance matrix. These mocks were produced through a hybrid combination of an approximate gravity solver and a reference simulation \citep{10.1093/mnras/stw248} that evolved $3840^3$ dark matter particles on a cubic volume of side $2.5$ Gpc/h, using the code {\tt GADGET-2} \citep{2005Natur.435..629S}, with a baseline cosmology given by $\{\Omega_b,\Omega_m,n_s,\sigma_8,h\}=\{0.0482,,0.307,0.961,0.829,0.6778\}$. Gravitationally-bound halos were identified using the Bound Density Maximum halo finder \citep{Klypin:1997sk}, which were subsequently populated with galaxies using the Halo Abundance Matching technique (HAM) \citep{Kravtsov_2004}, an alternative to the HOD method described above. The {\tt PATCHY} mocks were finally shaped into the realistic survey geometry of the BOSS CMASS dataset, also split into the separate NGC and SGC observed parts of the sky, while each galaxy was assigned a set of systematic weights (similarly to Eq. \eqref{eq:wc})
\begin{equation}\label{eq:wcPat}
w_{\rm c}(\bold{r}) = w_{\rm fc}(\bold{r})w_{\rm veto}(\bold{r}).
\end{equation}
The weights include fiber collisions, $w_{\rm fc}$, and a veto mask, $w_{\rm veto}$, capturing the rest of the associated shortcomings of the dataset. The galaxies are also assigned FKP weights, according to Eq. \eqref{eq:fkpweight}. Since the {\tt PATCHY} mocks were cast into a survey format, we treat them as the data and repeat the exact same procedure detailed in \S\ref{sec:dataset} in order to generate the resulting FKP field from Eq. \eqref{fkpfield}, but using the weighting scheme in Eq. \eqref{eq:wcPat} rather than in Eq. \eqref{eq:wc}. 

We will assume a cosmology-independent covariance matrix \citep{Kodwani_2019,refId0Car} and combine the {\tt PATCHY} mocks with the {\tt ABACUSSUMMIT} suite, even though strictly speaking the two sets correspond to different fiducial cosmologies and used different mock-generating procedures. We also note that mixing different ways of modeling the theory vector and the covariance matrix is common practice in BOSS analyses (eg. \citep{10.1093/mnras/stw1096,10.1093/mnras/stw3298,Ivanov_2020,d_Amico_2020,Philcox_2020,PhysRevD.105.043517}). Under this approximation, we use the {\tt ABACUSSUMMIT} fiducial cosmology (rather than the one of {\tt PATCHY}) to convert the galaxy positions from sky coordinates RA, DEC, and z into comoving Cartesian ones. We finally add that the {\tt PATCHY} mocks are accompanied by their own set of randomly generated catalogues, both for the NGC and the SGC, each containing $\sim 50 \times$ the number of objects in the corresponding actual galaxy mock.

%%%%%%%%%%%%%%%%%%%%%%%%%%%%%%%%%%%%%%%%%%%%%%
\subsection{Power spectrum}\label{sec:Pkeval}

In this section we explain how we evaluate the multipoles of the anisotropic redshift-space power spectrum, the predictions of which we will use as a reference to assess the performance of the WST. Given that the BOSS data (as well as the {\tt PATCHY} mocks) and the {\tt ABACUSSUMMIT} mocks come in a different format, we will follow a different strategy to extract the power spectrum multipoles in each case. 
%%%%%%%%%%%%%%%%%%%%%%%%%%%%%%%%%%%%%%%%%%%%%%

\subsubsection{BOSS data $\&$ {\tt PATCHY} mocks}\label{sec:dataPk}

When working with galaxy data or mocks in a non-trivial survey geometry, with a corresponding FKP field from Eq. \eqref{fkpfield}, the multipoles of the anisotropic redshift-space power spectrum, $\hat{P}_\ell(k)$, can be evaluated using the `Yamamoto' estimator \citep{1994ApJ...426...23F,10.1093/pasj/58.1.93,10.1093/mnrasl/slv090,PhysRevD.92.083532}:
\begin{align} \label{eq:Yamamoto}
&\hat{P}_\ell(k) = \frac{(2\ell +1)}{I_2} \int \frac{d \Omega_k}{4 \pi} \left[ \int d^3 \bold{r_1} F(\bold{r_1}) e^{i \bold{k} \cdot \bold{r_1}} \right. \\
& \left. \times \int d^3 \bold{r_2} F(\bold{r_2}) e^{i \bold{k} \cdot \bold{r_2}} L_\ell(\hat{\bold{k}} \cdot \hat{\bold{r_2}}) - P_\ell^{\rm shot} (\bold{k})\right]\nonumber.
\end{align}
Here $L_\ell$ is the Legendre polynomial of order $\ell$, $d \Omega_k$ is the differential solid angle element in Fourier space and the term $P_\ell^{\rm shot}$ is the shot noise contribution:
\begin{equation}\label{eq:YamamotoShot}
P_\ell^{\rm shot}(\bold{k}) = (1+\alpha_r) \int d^3\bold{r} w^2(\bold{r})\bar{n}_g(\bold{r}) L_\ell(\hat{\bold{k}} \cdot \hat{\bold{r}}),
\end{equation}
which vanishes for higher order multipoles $\ell>0$, and where $w(\bold{r})=w_c(\bold{r}) w_{\rm FKP}(\bold{r})$. We use Eqs. \eqref{eq:Yamamoto} and \eqref{eq:YamamotoShot} to extract the galaxy power spectrum multipoles from the BOSS catalogues and the {\tt PATCHY} mocks, both of which are shaped into a realistic survey geometry. We do so using {\tt nbodykit}, which follows the optimized Fast Fourier Transform (FFT)-based implementation of Eq. \eqref{eq:Yamamoto}, developed by Ref. \citep{Hand_2017}. In particular, we use Eq. \eqref{fkpfield} to evaluate the FKP field using the Triangular Shaped Cloud (TSC) mass assignment scheme \citep{hockney1981computer} on NGRID$=500$ cubic cells on the side, which corresponds to a high enough resolution for an accurate description of the scales we will consider. We then evaluate the first 2 non-vanishing multipoles, $\ell=\{0,2\}$, of the power spectrum through Eq. \eqref{eq:Yamamoto} using 46 linearly spaced bins of width $\Delta k = 0.01$ h/Mpc within the $k$ range $0.001-0.50$ h/Mpc. Our chosen $\Delta k$ value has been found to be adequate for recovering the information encoded in the power spectrum \citep{PhysRevD.102.063504}. We finally discard wavenumbers larger than $k_{\rm max}=0.25$ h/Mpc, a choice that both matches the ones of previous BOSS analyses \citep{10.1093/mnras/stw1096,10.1093/mnras/stw3298,Ivanov_2020,d_Amico_2020,Philcox_2020,PhysRevD.105.043517} and also guarantees that the power spectrum and WST both reach a similar minimum scale, such that we perform a fair comparison between the two estimators.

%%%%%%%%%%%%%%%%%%%%%%%%%%%%%%%%%%%%%%%%%%%%%%
\subsubsection{{\tt ABACUSSUMMIT} mocks}\label{sec:AbacPk}

The {\tt ABACUSSUMMIT} mocks we work with were produced, by design, in the regular {\tt ABACUS} periodic cubic geometry with $L_{\rm Box}=2000$ Mpc/h, with the redshift-space anisotropy applied along their Cartesian $\hat{\bold{z}}$-axis and without any systematic weights applied to them. As a result, their corresponding power spectrum multipoles can be straightforwardly evaluated using standard FFT-based algorithms on a periodic box, which are also supported by {\tt nbodykit}. We do so using a TSC mass assignment scheme with NGRID$=1700$ cells on the side and the same $k$-binning strategy adopted for the data in \S\ref{sec:dataPk}, a set of choices that was once again confirmed to guarantee sufficient accuracy for the range of scales included in our analysis.

However, we should be careful when comparing power spectra evaluated from data on a survey geometry against predictions obtained from periodic cubic mocks. This is because of the fact that, if $\delta(\bold{x})$ is the regular galaxy density field obtained from a periodic volume, and $W(\bold{x})$ the survey window function (that includes both the survey geometry and systematic weights), then a survey will observe a masked density field, $\delta'(\bold{x})$, given by \citep{1994ApJ...426...23F,10.1093/mnras/stw2576}:
\begin{equation}\label{eq:deltamasked}
\delta'(\bold{x}) = \delta(\bold{x})W(\bold{x}),
\end{equation}
which corresponds to a convolution in Fourier space. The FKP field of Eq. \eqref{fkpfield} is such an example. As a consequence, if $P(\bold{k})$ is the theory power spectrum corresponding to $\delta(\bold{x})$ (estimated, for example, from the {\tt ABACUSSUMMIT} mocks or a perturbation theory model) then the power spectrum, $\hat{P}(\bold{k})$, observed from a survey with non-trivial geometry will be modified as follows \citep{1994ApJ...426...23F,10.1093/mnras/stw1096,10.1093/mnras/stu1051,10.1093/mnras/stw2576,10.1093/mnras/stw3298}:
\begin{equation}\label{eq:Pkmasked}
\hat{P}(\bold{k}) = \int \frac{d^3 \bold{k}'}{\left(2 \pi\right)^3} P(\bold{k}) |\tilde{W}(\bold{k}-\bold{k}')|^2,
\end{equation}
where
\begin{equation}\label{eq:WFT}
\tilde{W}(\bold{k}) = \frac{\alpha_r}{I_2^{1/2}} \int d^3 \bold{r} \bar{n}_s (\bold{r}) e^{i \bold{k}\cdot \bold{r}},
\end{equation}
is the Fourier Transform (FT) of the window function, that obeys the normalization
\begin{equation}\label{eq:WFTnorm}
\int \frac{d^3 \bold{k}}{\left(2 \pi\right)^3} |\tilde{W}(\bold{k})|^2 = 1.
\end{equation}
To evaluate the multipoles, $\hat{P}_\ell(k)$ in Eq. \eqref{eq:Pkmasked}, we transform the {\tt ABACUS}-derived power spectrum multipoles, $P_\ell(k)$, to get the corresponding correlation function multipoles, $\xi_{\ell}(s)$, as follows \citep{10.1093/mnras/stw2576,10.1093/mnras/stw3298}\footnote{We use the package {\tt mcfit}, which is publicly available at \url{https://github.com/eelregit/mcfit} and implements the FFTLog algorithm \citep{10.1046/j.1365-8711.2000.03071.x}.}:
\begin{equation}\label{eq:xiell}
\xi_{\ell}(s) = i^{\ell} \int \frac{dk k^2}{2 \pi^2} j_{\ell}(k s) P_\ell(k),
\end{equation} 
with $j_{\ell}(k s)$ the spherical Bessel functions of order $\ell$. Given the configuration space multipoles of Eq. \eqref{eq:WFT}, $W^2_{\ell}(s)$, we can modify the correlation function multipoles to account for this effect. Explicitly, the window-corrected multipoles, $\hat{\xi}_{\ell}(s)$, will be given (up to $\ell=2$) by \citep{10.1093/mnras/stw3298}:
\begin{align} \label{eq:xiellconv}
\hat{\xi}_0(s) &= \xi_0(s) W^2_0(s) + \frac{1}{5}\xi_2(s) W^2_2(s), \nonumber \\
\hat{\xi}_2(s) &= \xi_0(s) W^2_2(s) + \xi_2(s) \left[W^2_0(s) + \frac{2}{7} W^2_2(s)\right].
\end{align}
Finally, these can be transformed back to the Fourier space, in order to give the window-corrected multipoles, $\hat{P}_\ell(k)$, through:
\begin{equation}\label{eq:Pkellmasked}
\hat{P}_\ell(k) = (-i)^{\ell} 4\pi \int ds s^2 j_{\ell}(ks)\hat{\xi}_{\ell}(s).
\end{equation}

We use Eq. \eqref{eq:Pkellmasked} to evaluate the window-corrected power spectrum multipoles from the {\tt ABACUSSUMMIT} mocks, to be compared against the corresponding results from the BOSS data and the {\tt PATCHY} mocks\footnote{We use the corrected normalization coefficients for the power spectrum, as explained in detail in Ref. \citep{Beutler_2021}.}. To apply Eq. \eqref{eq:xiellconv}, we use the publicly available results for $W^2_{\ell}(s)$ provided by Ref. \citep{10.1093/mnras/stw3298} (which evaluates these functions with the pair-counting method proposed by Ref. \citep{10.1093/mnras/stw2576}), for both the NGC and the SGC patches of the BOSS CMASS sample, separately. Following the standard practice in previous BOSS analyses \citep{10.1093/mnras/stw1096,10.1093/mnras/stw3298,10.1093/mnras/stu1051}, we average over the window contributions for the North and the South, according to Eq. \eqref{eq:NplusS}. An alternative way to handle the effects of survey geometry would be to de-convolve the data, instead, as more recently proposed by Ref. \citep{Beutler_2021}.

Furthermore, as we discussed in \S\ref{sec:dataset}, the assumption of a given (and potentially incorrect) cosmology when converting the data (and the {\tt PATCHY} mocks) from sky coordinates, RA, DEC and z, into comoving coordinates might introduce an error in our analysis \citep{1979Natur.281..358A}. To account for this AP effect on the estimation of the power spectra from the mocks, we work as follows \citep{2022MNRAS.509.1779L}: if $r_{\parallel}$ and $r_{\perp}$ are the mock galaxy coordinates parallel and perpendicular to the line of sight (which coincides with the Cartesian $\hat{z}$ direction for the cubic mocks), they should then be re-scaled according to the following relations: 
\begin{equation}\label{eq:rparAP}
r_{\parallel, \rm ref} = r_{\parallel, \rm sim} \left(\frac{H_{\rm sim}(z) H_{0,\rm ref}}{H_{0,\rm sim}H_{\rm ref}(z)} \right),
\end{equation}
and
\begin{equation}\label{eq:rperpAP}
r_{\perp, \rm ref} = r_{\perp, \rm sim} \frac{d_{A,\rm ref}(z)}{d_{A,\rm sim}(z)}.
\end{equation}
The subscripts `sim' and `ref'  in Eqs. \eqref{eq:rparAP} and \eqref{eq:rperpAP} denote the true cosmology of each mock and the reference cosmology assumed for the conversion (corresponding to the fiducial, in our case), respectively, and $d_A$ is the comoving angular diameter distance. That is, before we apply Eqs. \eqref{eq:deltamasked}-\eqref{eq:Pkellmasked} to evaluate the power spectrum multipoles, the galaxy coordinates of each mock are first re-scaled according to Eqs. \eqref{eq:rparAP} and \eqref{eq:rperpAP}. Alternatively, one could use an analytical prediction of the AP effect on the anisotropic power spectra (as, for example, done in \citep{Ivanov_2020,d_Amico_2020,Philcox_2020,PhysRevD.105.043517,Chen_2022,Zhang_2022}), a procedure that is equivalent to the one described above. 

\begin{figure}[ht]
\includegraphics[width=0.49\textwidth]{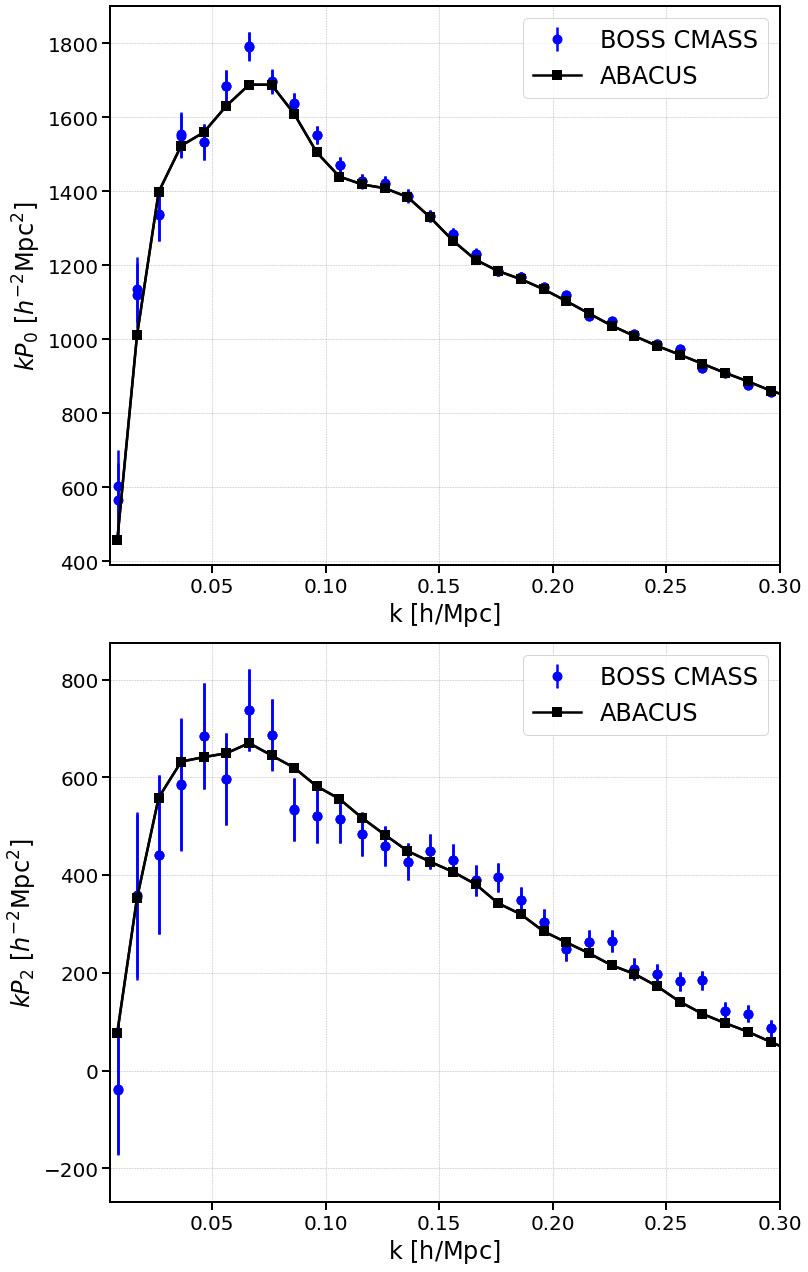}
\caption{Redshift-space monopole (top panel) and quadrupole (bottom panel) of the galaxy power spectrum evaluated from the BOSS CMASS dataset (blue circles) and the {\tt ABACUSSUMMIT} mocks (black squares) for the fiducial cosmology. The $1\sigma$ error bars on the data have been evaluated from the 2048 {\tt PATCHY} mock realizations.}
\label{fig:Pk} 
\end{figure}

Finally, in Fig.~\ref{fig:Pk} we show the power spectrum multipoles for the base cosmology, as obtained from the {\tt ABACUSSUMMIT} mocks using Eq. \eqref{eq:Pkellmasked} and for the BOSS CMASS dataset, as well as the {\tt PATCHY} mocks, using Eq. \eqref{eq:Yamamoto}. We highlight the excellent monopole agreement between the {\tt ABACUSSUMMIT} prediction and the corresponding one from the data, down to scales smaller than the ones we work with in this analysis. 

%%%%%%%%%%%%%%%%%%%%%%%%%%%%%%%%%%%%%%%%%%%%%%
\subsection{WST}\label{sec:WSTeval}

To evaluate the WST coefficients for all datasets and mocks used in our analysis, we make use of the publicly available package {\tt KYMATIO} \citep{2018arXiv181211214A}\footnote{\url{https://www.kymat.io/}.}, which implements the WST Eqs. \eqref{eq:WSTcoeff:sol} for an input 3D density field $I(\bold{x})$\footnote{Strictly speaking, {\tt KYMATIO} evaluates the sum over all pixels of the input field, rather than the mean, which is the same up to a normalization, and thus equivalent for parameter inference applications.}. We evaluate all input density fields on a cubic box of side $L_{\rm box}=2820$ Mpc/h, with NGRID$=282$ grids on the side and the TSC mass assignment scheme, while always adopting the choices $J=4, L=4, \sigma=0.8$ and $q=0.8$. The choice of a grid cell with a side equal to $10$ Mpc/h guarantees that the WST uses information from a minimum scale that is both similar to the one used for the power spectrum multipoles (for which $k_{max}=0.25$ h/Mpc) and also ensures that we do not extract information from a regime that would make the evaluation susceptible to small-scale systematics. This combination corresponds to a basis of $S_0+S_1+S_2 = 76$ total WST coefficients (from Eq. \eqref{Stotal}).
%%%%%%%%%%%%%%%%%%%%%%%%%%%%%%%%%%%%%%%%%%%%%%

\subsubsection{BOSS data $\&$ {\tt PATCHY} mocks}\label{sec:dataWST}

In order to extract the WST coefficients from the data and the {\tt PATCHY} mocks, we need to apply Eq. \eqref{eq:WSTcoeff:sol} with the corresponding FKP field evaluated from Eq. \eqref{fkpfield}, as input. However, we need to be careful at this point, because Eq.~\eqref{eq:WSTcoeff:sol} assumes a periodic 3D cube as input, rather than a masked density field such as the FKP one. To overcome this obstacle, we modified the public version of {\tt KYMATIO} such that Eq. \eqref{eq:WSTcoeff:sol}, and in particular its fundamental operations of wavelet convolution and modulus, can be used with a masked density field of the form \eqref{fkpfield} (or more generally \eqref{eq:deltamasked}) as input (with the technical details discussed in Appendix \S\ref{sec:App_WSTmask}). We finally proceed to extract the WST coefficients from the CMASS dataset, as well as from the 2048 {\tt PATCHY} mocks, using this modified version.
%%%%%%%%%%%%%%%%%%%%%%%%%%%%%%%%%%%%%%%%%%%%%%
\subsubsection{{\tt ABACUSSUMMIT} mocks}

As we already discussed in \S\ref{sec:AbacPk}, the {\tt ABACUS}-derived mocks from {\tt ABACUSHOD} have a 3D cubic geometry. However, as we discussed in \S\ref{sec:dataWST}, the WST coefficients extracted from the data were computed from the masked density field of Eq. \eqref{fkpfield}. This implies that, just like in the case of the power spectrum, the effect of survey geometry needs to be taken into consideration in our WST predictions from {\tt ABACUS} (in fact, the WST analysis is even more sensitive to the survey geometry). In the absence of a model to apply this window correction on the evaluated statistic directly, such as Eq. \eqref{eq:Pkellmasked} for the power spectrum, we proceed to cut the {\tt ABACUS} cubes into the shape of the BOSS CMASS data\footnote{Alternatively, inpainting techniques could be considered \citep{Mohammad:2021ylk}.}. For this,  we use the public code {\tt make\_survey} \citep{White:2013psd} \footnote{Available at \url{https://github.com/mockFactory/make_survey}}, with the exact CMASS angular footprint (for each one of the NGC and the SGC patches) and redshift range, as input. Before feeding the cubic mocks into {\tt make\_survey}, we first undo the redshift-space distortions effect originally applied along the Cartesian $\hat{\bold{z}}$-axis, such that the code can then implement it along the sky radial direction, resembling the realistic configuration of the actual survey. The necessary galaxy velocities for these RSD manipulations are also provided upon the mock generation by {\tt ABACUSHOD}. The final resulting output is an equivalent set of galaxy mocks in sky coordinates RA, DEC, and z that exactly match the 3D geometry of the observed CMASS dataset. To make sure that the clustering properties were not affected during the cut sky implementation, we evaluate the power spectrum multipoles from the new re-shaped mocks from Eq. \eqref{eq:Yamamoto} (using, also, an additional set of randoms produced with the same procedure) and confirm that the result matches the one from the corresponding cubic box using Eq. \eqref{eq:Pkellmasked}, for the fiducial cosmology. One could actually use either of those ways of evaluating the power spectrum, since they are equivalent when handled properly, and which choice to go with is ultimately a matter of preference. Having confirmed the robustness of this procedure, we then proceed to evaluate the FKP fields from all the cut {\tt ABACUS} mocks using Eq. \eqref{fkpfield} (but with all the weights set to 1, which we then feed into the set in Eq. \eqref{eq:WSTcoeff:sol} so as to finally get the WST coefficients for all cosmologies of Table~\ref{table:1}.

Lastly, as with the power spectrum analysis laid out in \S\ref{sec:Pkeval}, we need to also account for the AP effect introduced by the assumption of a given cosmology when converting the data into comoving coordinates. In the WST case, the mock data were converted (using the true cosmology of each simulation) into sky coordinates upon the cut-sky procedure we described above. To account for the AP effect, we then use the reference fiducial cosmology in order to convert these data back into comoving coordinates, a step that is necessary to evaluate the FKP field from Eq. \eqref{fkpfield}, out of which the WST coefficients will be extracted. This step serves as the WST equivalent to the AP re-scalings applied in Eqs. \eqref{eq:rparAP} and \eqref{eq:rperpAP}  for the power spectrum multipoles. 

\begin{figure}[ht]
\includegraphics[width=0.48\textwidth]{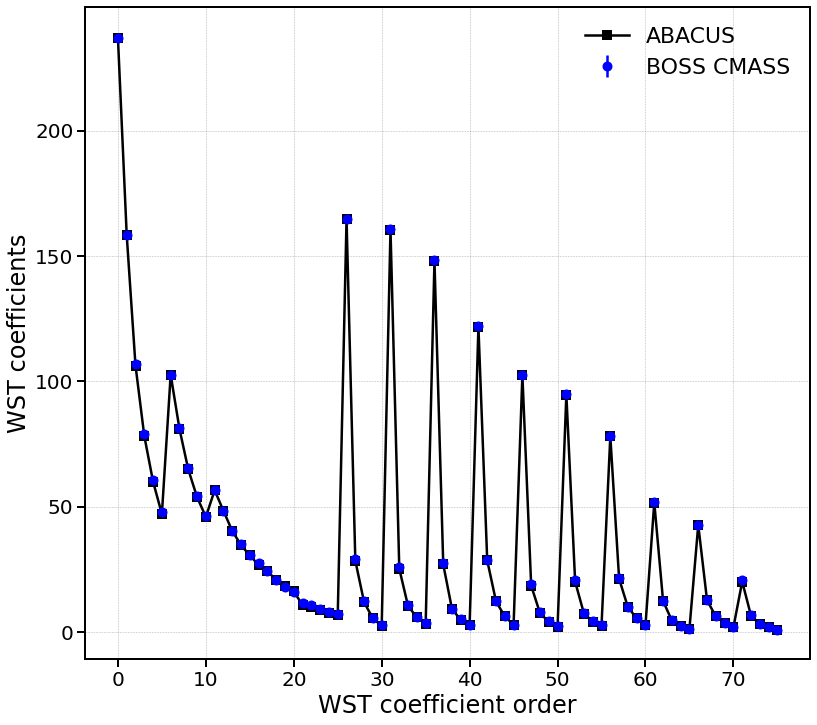}
\caption{All 76 WST coefficients evaluated from the BOSS CMASS dataset (blue circles) and the {\tt ABACUSSUMMIT} mocks (black squares) for the fiducial cosmology. The WST coefficients populate the data vector in order of increasing values of the $j_1$ and $l_1$ indices, with the $l_1$ index varied faster. The $1\sigma$ error bars on the data (which are too small to be clearly seen on the plot) have been evaluated from the 2048 {\tt PATCHY} mock realizations.}
\label{fig:WST} 
\end{figure}

In Fig.~\ref{fig:WST}, we plot the WST coefficients evaluated from the {\tt ABACUSSUMMIT} mocks, together with the corresponding results from the CMASS data, for the fiducial cosmology. The level of agreement between the {\tt ABACUS} model and the data, which is even better than the one in the power spectrum case, confirms the validity of our model constructed for the WST.

%%%%%%%%%%%%%%%%%%%%%%%%%%%%%%%%%%%%%%%%%%%%%%
\subsection{Likelihood analysis}\label{sec:mcmc}

\begin{figure}[b]
\centering 
\includegraphics[width=0.49\textwidth]{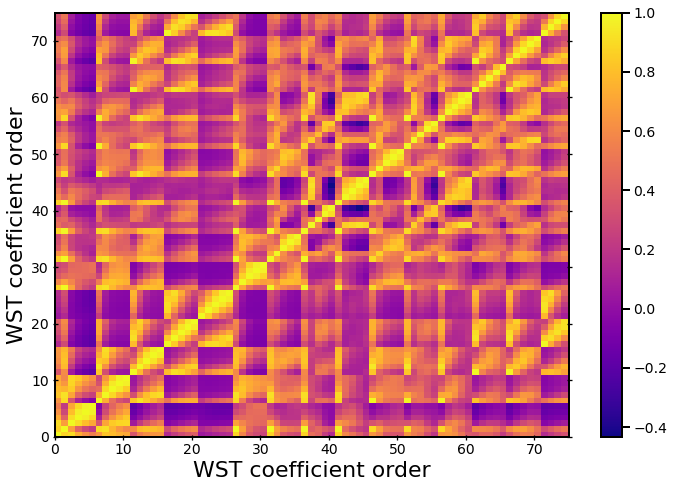}
\caption{Correlation matrix of all 76 coefficients for the WST evaluated at the fiducial cosmology. The WST coefficients populate the data vector in order of increasing values of the $j_1$ and $l_1$ indices, with the $l_1$ index varied faster, as in Fig.~\ref{fig:WST}.}
\label{Fig:WSTcov}
\end{figure}

\begin{figure}[ht]
\includegraphics[width=0.49\textwidth]{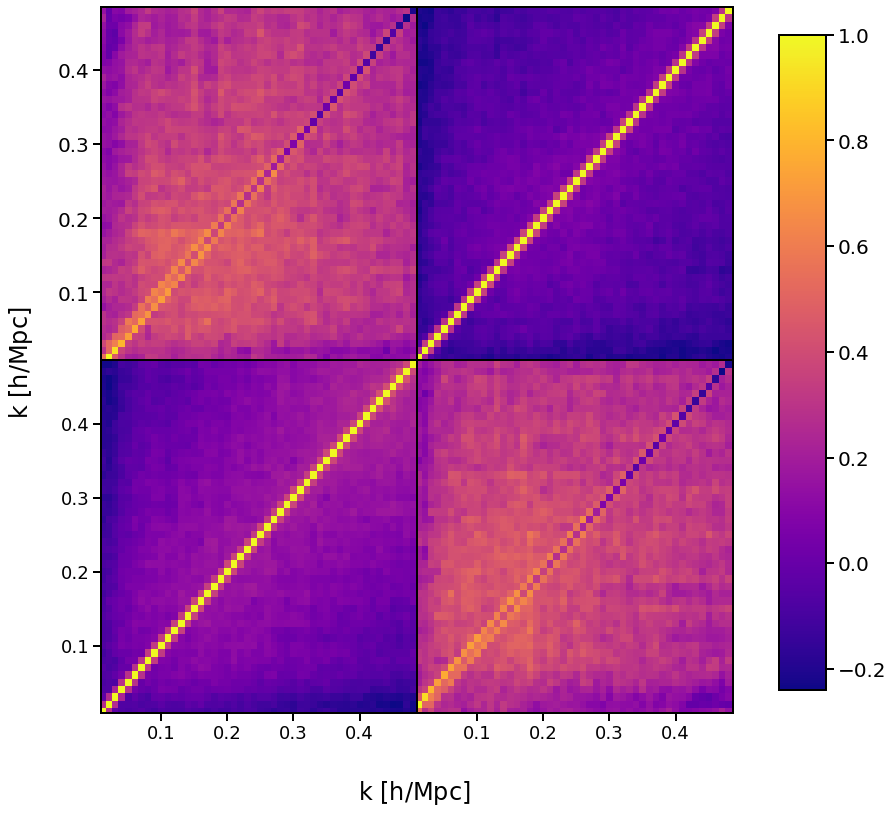}
\caption{Correlation matrix of the galaxy power spectrum multipoles, $l=\{0,2\}$, evaluated from the 2048 realizations of the {\tt PATCHY} mocks for the fiducial cosmology. In the $2\times2$ blocks, from bottom to top and from left to right, we visualize the auto and cross correlations of $\hat{P}_0$ and $\hat{P}_2$, respectively.}
\label{fig:Pkcov} 
\end{figure}

Having laid out the methodology on how to extract the clustering statistics from both the data and also from the two sets of mocks used in this work, we now proceed to explain our strategy for combining these necessary ingredients into a likelihood analysis of the BOSS dataset. In particular, if $\bold{X}$ is our target estimator (the WST coefficients or the power spectrum multipoles), we assume a Gaussian likelihood\footnote{The validity of this assumption was recently tested in the weak lensing application of Ref. \citep{Cheng:2021hdp}, in which the  probability distribution of the WST coefficients was found to be closer to Gaussian than that of the bispectrum. Also see Ref. \cite{Park:2022hzj} for an in-depth analysis of the validity of this assumption for various statistics.}, $\mathcal{L}(\theta|\bold{d})$, given by the following relation:
\begin{equation}\label{eq:LogL}
\log \mathcal{L}(\theta|\bold{d}) = -\frac{1}{2} \left[\bold{X}_\bold{d}-\bold{X}_t(\theta)\right]^{\rm T} C^{-1}\left[\bold{X}_{\bold{d}}-\bold{X}_t(\theta)\right] + {\rm const.},
\end{equation} 
where $\theta$ is the parameter we want to extract from the data ${\bf d}$ and $\bold{X}_\bold{d}$ is the value extracted from the BOSS data $\bold{d}$. The covariance matrix $C$ is estimated from the $N_{\rm mocks}=2048$ {\tt PATCHY} mocks:
\begin{equation}\label{eq:covmat}
C = \frac{1}{\rm N_{mocks} - 1}\sum_{\rm k=1}^{\rm N_{\rm mocks}} \left(\bold{X}^k_P-\bar{\bold{X}}_P\right)\left(\bold{X}^k_P-\bar{\bold{X}}_P\right)^{\rm T},
\end{equation} 
with $\bar{\bold{X}}_P$ the mean prediction from the $N_{\rm mocks}$. In order to de-bias our prediction for the inverse covariance matrix, $C^{-1}$, we apply the Hartlap correction factor \citep{refId22}, as follows:
\begin{equation}\label{Hartlap}
\hat{C}^{-1}= \frac{N_{\rm mocks}-N_{\bold{d}}-2}{N_{\rm mocks}-1}C^{-1}, 
\end{equation}
where $N_{\bold{d}}=58$ when working with the $l={0,2}$ multipoles of the galaxy power spectrum (up to $k_{\rm max}=0.25$ h/Mpc) and $N_{\bold{d}}=76$ for the WST coefficients as the data vector. Prior to inversion, we make sure that the covariance matrices for both estimators are well-conditioned and can thus be safely inverted in order to be used in the likelihood in Eq. \eqref{eq:LogL}. The convergence of the WST covariance is confirmed in Appendix \ref{sec:App_convergence}. The correlation matrix, $C_{ij}/(C_{ii}C_{jj})$, of the WST coefficients is shown in Fig. \ref{Fig:WSTcov}, while the corresponding matrix for the power spectrum multipoles is presented in Fig. \ref{fig:Pkcov}, both evaluated at the fiducial cosmology.

Crucially, the remaining quantity to determine in Eq. \eqref{eq:LogL} is the model, $\bold{X}_t(\theta)$, which captures the dependence of each estimator on the target set of cosmological parameters, $\theta$. In our case this is an 11-dimensional vector consisting of the 4 cosmological parameters varied in Table~\ref{table:1} and the 7 nuisance parameters of the HOD model, from Eq. \eqref{eq:HODfid}. We construct this model as follows: if $\theta_{\rm fid}$ is the vector of the parameter values determining our fiducial cosmology, given by the parameters in the first row of Table~\ref{table:1}, and the associated best-fit HOD parameters from Eq. \eqref{eq:HODfid}, we model the parameter dependence using the following expansion:
\begin{equation}\label{Xtaylor}
\bold{X}_t(\theta) = \bold{X}_t(\theta_{\rm fid}) + (\theta-\theta_{\rm fid})\nabla_\theta \bold{X},
\end{equation}
where the gradient $\nabla_\theta \bold{X}$ in Eq. \eqref{Xtaylor} is straightforwardly determined using the derivatives constructed from the derivative grid cosmologies of Table~\ref{table:1}, as we explained in \S\ref{sec:Mocksabac}. We have carefully checked and confirmed that the combination of the vector dimensionality, derivative step size and number of HOD realizations used is sufficient for the derivatives to be well-converged, for all parameters and for both estimators (see Appendix \S\ref{sec:App_compression} for details). It should be clarified, at this point, that being a first order expansion, Eq. \eqref{Xtaylor} needs to be evaluated using a fiducial cosmology $\theta_{\rm fid}$ sufficiently close to the true one, such that the derivative approximation only breaks down far away from the (true) maximum of the likelihood and the correct cosmology is recovered after a likelihood analysis of the data. Since our chosen $\theta_{\rm fid}$ corresponds to the mean values of the {\it Planck} 2018 \citep{Planck2018} $\Lambda$CDM cosmology, and as it can be also inferred by the very good agreement between the fiducial theory prediction and the one from the data shown in Figs. \ref{fig:Pk} and \ref{fig:WST}, this requirement is satisfied in our analysis. We do, however, point out that the since the Taylor expansion \eqref{Xtaylor} will inevitably break down far away from the fiducial part of the parameter space, our model will fail to capture any potential non-Gaussianities in the likelihood, for which a full model (e.g., an emulator) for $\bold{X}_t(\theta)$ would be necessary. We add that a very similar expansion was also recently used in the Dark Energy Survey (DES) year-1 data re-analysis of Ref. \citep{Hadzhiyska_2021}, which employed a hybrid combination of perturbation theory and {\tt ABACUSSUMMIT} simulations in order to model the lensing power spectrum needed for their analysis \footnote{We note that Ref. \citep{Hadzhiyska_2021} was able to correct the errors caused by such a Taylor expansion, using the HALOFIT model for the matter power spectrum. Given that no such possibility is available for the cosmological dependence of the WST coefficients, a correction of this kind was not possible in our case.}. We leave the construction of an actual WST emulator for future work. Finally, we clarify that, even though a wide variety of approaches exist for a more accurate modeling of the cosmological dependence of the galaxy power spectrum or correlation function (e.g., perturbation theory \citep{Ivanov_2020,d_Amico_2020,Philcox_2020,PhysRevD.105.043517,Chen_2022,Zhang_2022}) or emulators \citep{Nishimichi_2019,Miyatake:2020uhg,PhysRevD.102.063504,Zhai:2022yyk,Yuan:2022jqf}), we use the model from Eq. \eqref{Xtaylor} for the power spectrum, as well, in order to guarantee a fair comparison against the performance of the WST.

In order to perform a posterior analysis, we sample the likelihood from Eq. \eqref{Xtaylor} using the Markov Chain Monte Carlo (MCMC) sampler {\tt emcee}\citep{2013PASP..125..306F} \footnote{Publicly available in \url{https://emcee.readthedocs.io/en/stable/}}, choosing 100 walkers, 500 'burn-in' steps and 100000 steps for our main runs. We use flat unrestricted priors (with the walkers initialized in the range $0-1.05$) for all parameters, with the exception of $\omega_b$, for which we also consider the case of a Gaussian prior determined from Big Bang Nucleosynthesis (BBN) \citep{Planck2018} and the measurement of helium and deuterium primordial abundances \citep{Aver_2015,Cooke_2018,Sch_neberg_2019}:
\begin{equation}\label{omegabprior}
\omega_b = 0.02268 \pm 0.00038,
\end{equation}
a choice commonly adopted in analyses of BOSS data \citep{Ivanov_2020,d_Amico_2020,Philcox_2020}.
Finally, to check the convergence of our chains, we monitor the mean integrated autocorrelation time and make sure its value is at least 2 orders of magnitude lower than the total number of steps used, following the procedure laid out by Ref. \citep{2013PASP..125..306F}. We similarly also monitor the mean value of the acceptance fraction and make sure it always falls within the reasonable range of values of $0.3-0.5$.

%%%%%%%%%%%%%%%%%%%%%%%%%%%%%%%%%%%%%%%%%%%%%%
\section{Results}\label{sec:Results}

Before presenting our results from the likelihood analysis explained in \S\ref{sec:mcmc}, we briefly summarize the procedure we follow in order to determine the posterior of the Hubble constant, $h$, as a derived parameter.
As stated in \S\ref{sec:Mocksabac}, in the {\tt ABACUSSUMMIT} simulations the value of the angular size of the sound horizon at last scattering, $\theta_{\star}$, is kept fixed to the corresponding mean value derived from measurements by the {\it Planck} satellite, $100 \theta_{\star}=1.041533$. 
This implies that we can compute $h$ for each point in our chains and derive its posterior, which we will also discuss below as the fifth cosmological parameter determined from our analysis. 

\begin{figure}[ht!]
\includegraphics[width=0.49\textwidth]{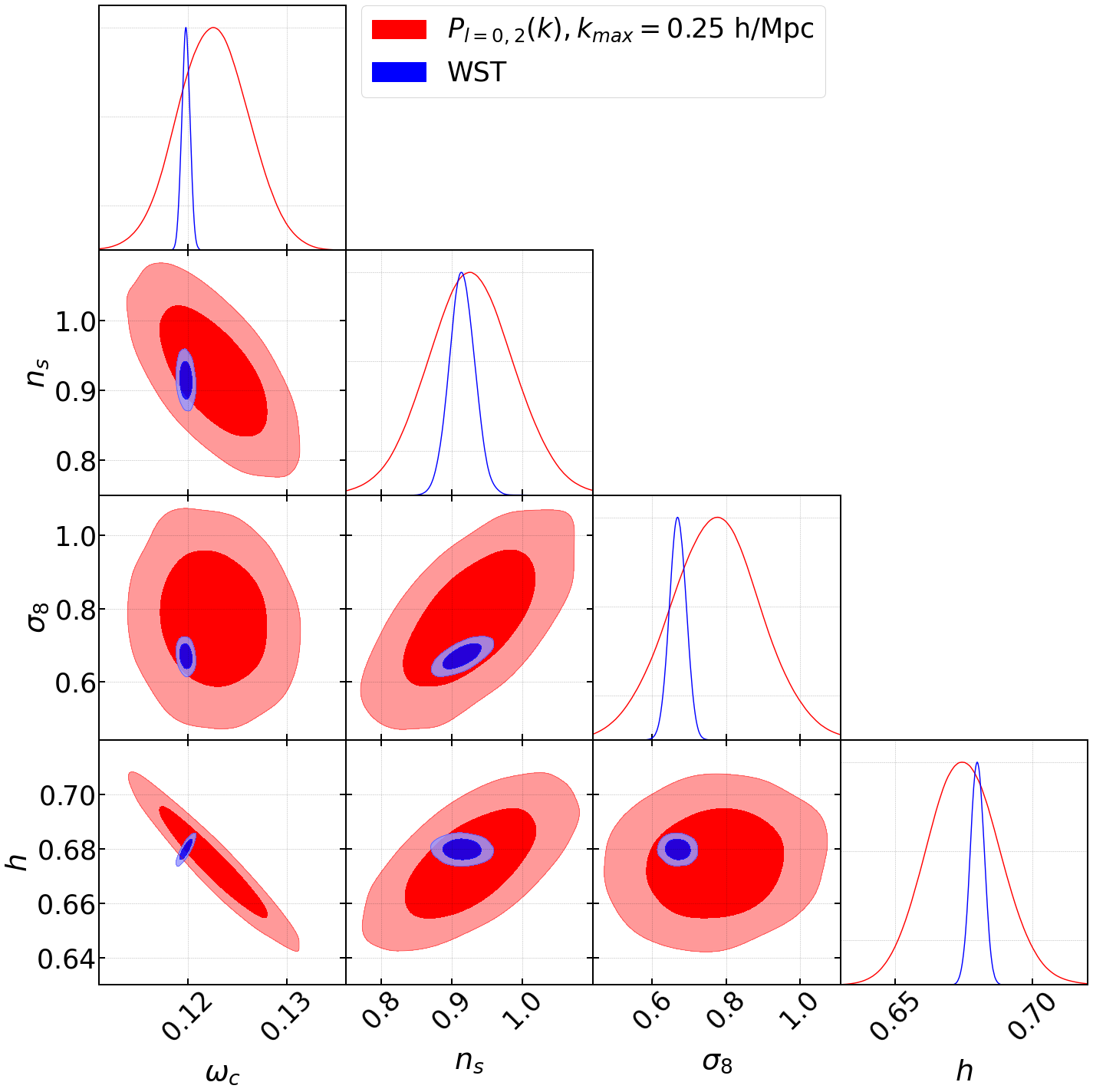}
\caption{Constraints on the cosmological parameters obtained from the combined monopole and quadrupole of the galaxy power spectrum evaluated up to $k_{\rm max}=0.25$ Mpc/h (red contours), as well as from the WST coefficients defined in \S\ref{sec:WSTeval} (blue contours). The results shown above were obtained after imposing a BBN Gaussian prior on the value of $\omega_b = 0.02268 \pm 0.00038$.}
\label{fig:BBNpriorOmegab_WST_vs_Pk}
\end{figure}

We now discuss the results of our analysis of the BOSS CMASS data using the WST and the power spectrum multipoles, starting with the case in which the BBN prior \eqref{omegabprior} was imposed on the value of $\omega_b$. In Fig.~\ref{fig:BBNpriorOmegab_WST_vs_Pk}, we show the resulting 2-dimensional posterior probability distribution function of the 3 other cosmological parameters explicitly varied in the likelihood \eqref{eq:LogL}, together with the derived Hubble constant $h$, all of which have been marginalized over the 7 nuisance parameters of the HOD model. In addition, the mean and $1\sigma$ error values obtained from the two estimators for all cosmological parameters are listed in Table \ref{table:2}, while the full corner plot is presented in Appendix \S\ref{sec:App}. 

We notice the relative consistency between the corresponding mean values for the parameters extracted from the two estimators, the differences of which never exceed the respective $1\sigma$ values from the power spectrum. Furthermore, all values are broadly consistent with the ones found from recent re-analyses of BOSS data \citep{Ivanov_2020,d_Amico_2020,Philcox_2020,PhysRevD.105.043517,Chen_2022,Zhang_2022}, a fact that confirms the robustness of the WST as a tool to be used for cosmological analyses. More importantly, in addition to being able to serve as a reliable clustering statistic to infer cosmological parameters from the LSS, the WST is found to deliver significant improvements in the inferred $1\sigma$ errors for all cosmological parameters, in the range $3-8\times$ tighter. This finding demonstrates the potential carried in the use of the WST as a way to access the non-Gaussian information encoded in the LSS data (as suggested in our previous work \citep{Valogiannis:2021chp}), and thus subsequently improve the errors obtained on cosmological parameters. 

\begin{figure}[ht!]
\includegraphics[width=0.49\textwidth]{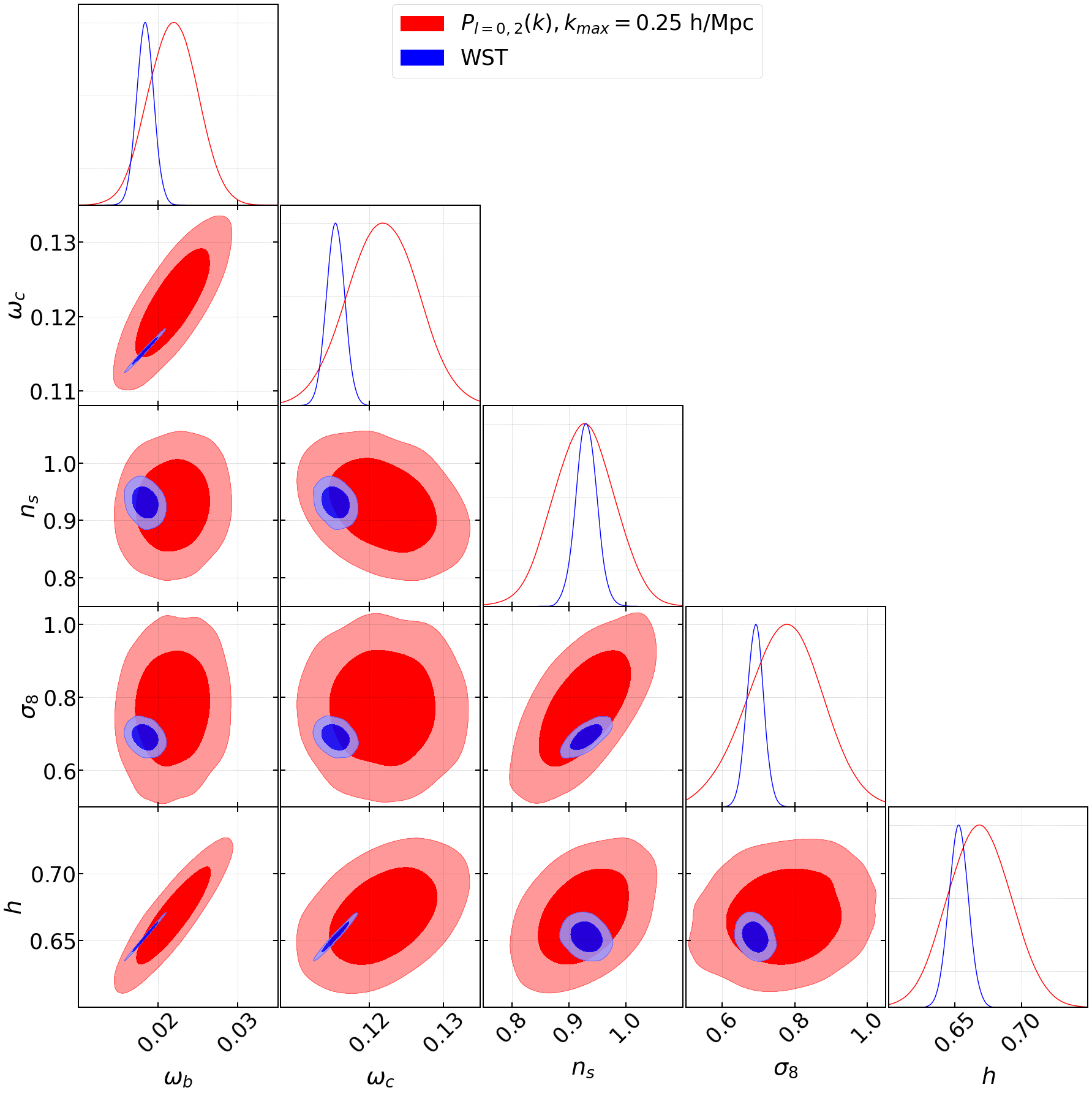}
\caption{Same as in Fig.~\ref{fig:BBNpriorOmegab_WST_vs_Pk}, but using a flat and uninformative prior on $\omega_b$.}
\label{fig:flatpriorOmegab_WST_vs_Pk} 
\end{figure}

In Fig.~\ref{fig:flatpriorOmegab_WST_vs_Pk} we show a case with no priors on any of the cosmological parameters (the results are also summarized in Table \ref{table:2}). We observe a similar trend as in the previous case, with the WST once again outperforming the regular power spectrum with respect to the obtained $1\sigma$ errors, by a factor of $3-5$. The inferred mean values of the 5 cosmological parameters are once again consistent, within $1\sigma$, between the two statistics, despite the fact that the totally unrestricted priors led to a relatively lower value of $\omega_b$ (relative to the BBN prior), and subsequently of $h$ (through the fixed $\theta_{\star}$). 

We also briefly comment on the results obtained from a likelihood analysis using the vector of WST coefficients up to first order, only. In this case, the inferred $1\sigma$ errors from the WST are improved compared to the corresponding power spectrum results by a factor of $1.1$ to $2.0$. 

A fair comparison with previous analyses of the BOSS data \citep{Ivanov_2020,d_Amico_2020,Philcox_2020,PhysRevD.105.043517,Chen_2022,Zhang_2022,2022arXiv220410392C,2022arXiv220411868B} is difficult given several differences between those works and ours, the major ones being the approximate model we employed from Eq. \eqref{Xtaylor} and the fact that the Hubble constant was not explicitly varied in our analysis, but was rather derived from a fixed $\theta_{\star}$\footnote{We also worked with a subset of the CMASS sample, rather than with the full CMASS and LOWZ samples, due to the limitations imposed by the HOD procedure.}. This is also the case when attempting to compare against recent BOSS analyses using emulators of the redshift-space power spectrum \citep{2021arXiv211006969K} or the correlation function \citep{Zhai:2022yyk,Yuan:2022jqf}, with the additional difference that the correlation function analyses focused on much smaller scales than the ones we worked with. It should be also noted that the 1$\sigma$ errors reported above are purely statistical, since we did not attempt to quantify how the various systematics and approximations adopted by our analysis would affect the final results. 

Furthermore, we clarify that in our WST analysis we have purely worked with unreconstructed density fields. In the context of traditional BOSS analyses, reconstruction algorithms \citep{2007ApJ...664..675E} have been shown to enable a more precise determination of the Baryon Acoustic Oscillation (BAO) peak position, the complementary information of which can further improve the constraints obtained using the full shape of the power spectrum \citep{Philcox_2020,Chen_2022} and also the bispectrum \citep{PhysRevD.105.043517}. Whether a WST analysis applied to the reconstructed density field can improve the constraints extracted on cosmological parameters is an interesting question, that we plan to explore in the future.
Putting these differences aside, we do highlight that the relative $1\sigma$ error from the power spectrum obtained on the Hubble constant is found to be equal to $2\%$, when using the BBN prior on $\omega_b$, a value that is similar to the one recently found by Refs. \citep{Ivanov_2020,Philcox_2020,PhysRevD.105.043517,Chen_2022}.

\begin{table}[h!]
\centering
\begin{tabular}{|p{0.3cm}||p{4.33cm}||p{3.9cm}|}
\hline 
& BBN prior on $\omega_b$& unrestricted priors \\
\end{tabular}

\centering
\begin{tabular}{|p{0.3cm}||p{2.1cm}|p{2.1cm}||p{1.8cm}|p{1.95cm}|}

\hline 
 & P(k)& WST&P(k)&WST\\
\hline 
$\omega_b$ & $0.02268^{+0.00036}_{-0.00036}$&$0.02225^{+0.00034}_{-0.00034}$&$0.0217^{+0.0043}_{-0.0043}$&$0.0184^{+0.0011}_{-0.0011}$\\
\hline
$\omega_c$ &$0.1225^{+0.0037}_{-0.0037}$&$0.120^{+0.00041}_{-0.00041}$&$0.1217^{+0.0058}_{-0.0058}$&$0.1154^{+0.0012}_{-0.0012}$\\
\hline
$n_s$ &$0.927^{+0.063}_{-0.063}$&$0.914^{+0.018}_{-0.018}$&$0.921^{+0.057}_{-0.049}$&$0.931^{+0.018}_{-0.018}$\\
\hline
$\sigma_8$ &$0.77^{+0.13}_{-0.13}$&$0.67^{+0.023}_{-0.023}$&$0.762^{+0.11}_{-0.094}$&$0.691^{+0.023}_{-0.023}$\\
\hline
$h$ & $0.675^{+0.014}_{-0.014}$&$ 0.68^{+0.0025}_{-0.0025}$&$0.668^{+0.024}_{-0.024}$&$0.653^{+0.0074}_{-0.0074}$\\
\hline
\end{tabular}
\caption{Mean values and $68\%$ confidence intervals for all cosmological parameters resulting from the posterior analysis of the power spectrum multipoles and the WST coefficients in the case of a BBN prior applied on the value of $\omega_b$ (left half), and the case of unrestricted priors (right half). All results are presented in the format `${\rm mean}^{+1 \sigma}_{-1 \sigma}$', after marginalization over all HOD parameters.}
\label{table:2}
\end{table}

Finally, we finish this section by commenting on the fact that the mean value of $\sigma_8$ obtained from our WST application is in very good agreement with results from recent BOSS re-analyses, which are also found to be in tension with the corresponding {\it Planck} value, in particular for the case of an unrestricted prior on $n_s$ \citep{PhysRevD.105.043517}. Imposing a {\it Planck} prior on $n_s$ somewhat raises the recovered mean value of $\sigma_8$ \citep{PhysRevD.105.043517,Chen_2022}, but is not large enough to completely alleviate the tension. Furthermore, cross-correlating BOSS clustering data with CMB lensing measured by {\it Planck} has been recently found to further lower the inferred value of $\sigma_8$ \citep{2022arXiv220410392C}. 

%%%%%%%%%%%%%%%%%%%%%%%%%%%%%%%%%%%%%%%%%%%%%%
\section{Conclusions}\label{sec:Conclusions}

In this work, we present the first application of the wavelet scattering transform on actual galaxy observations, through a WST analysis of the BOSS DR12 CMASS dataset.

Building upon our previous LSS application to 3D matter overdensity fields \citep{Valogiannis:2021chp}, we lay out the detailed methodology to capture additional layers of realism that are necessary to analyze galaxy observations obtained from a spectroscopic survey, such as BOSS. After capturing the effects of redshift-space anisotropy, non-trivial survey geometry, the shortcomings of the dataset through a set of systematic weights and the Alcock-Paczynski effect, we show how to transform a galaxy sample from redshift-space sky coordinates into the weighted Feldman-Kaiser-Peacock (FKP) field, which serve as the input of a WST scattering network. The resulting WST coefficients can then be treated as a well-defined basis that reflects the clustering properties of the observed sample, which we use as the main object of our BOSS analysis.

In order to model the cosmological dependence of the WST coefficients we use the state-of-the-art suite of {\tt ABACUSSUMMIT} simulations \citep{10.1093/mnras/stab2484}. These span the cosmological parameter space around the {\it Planck} 2018 $\Lambda$CDM cosmology \citep{Planck2018}, and have been HOD-tuned to match small-scale redshift-space correlation function of the BOSS CMASS sample in the redshift range $0.46<z<0.60$. For the evaluation of the WST covariance matrix, which is also necessary in addition to the model, we employ the publicly available {\tt MULTIDARK-PATCHY} mocks \citep{10.1093/mnras/stv2826,10.1093/mnras/stw1014}. We take all necessary steps to ensure that our mock theory predictions satisfy the same level of realism as the observations we compare them against, and also evaluate the multipoles of the anisotropic galaxy power spectrum, which we use as a benchmark to assess the performance of the WST. 

We then use our model to perform a likelihood analysis of the CMASS observations with the WST coefficients and the power spectrum multipoles. We obtain the posterior probability distributions of the 4 target cosmological parameters, $\{\omega_b,\omega_c,n_s,\sigma_8\}$, as well as the Hubble parameter, $h$, derived from the fixed value $\theta_{\star}$, all of which were marginalized over the 7 nuisance parameters of the HOD model. The analysis reveals a substantial improvement in the values of the $1\sigma$ errors predicted by the WST, which are tighter than the corresponding ones from the regular power spectrum by a factor of $3-8$, when a BBN prior is applied on the value of $\omega_b$, and by a factor in the range $3-5$ in the case of flat and uninformative priors for all of the parameters. At the same time, the inferred mean values of all cosmological parameters by the WST (as well as the power spectrum) are always found to be broadly consistent with the ones found by recent re-analyses of BOSS data \citep{Ivanov_2020,d_Amico_2020,Philcox_2020,PhysRevD.105.043517,Chen_2022,Zhang_2022,2022arXiv220410392C}, demonstrating, overall, that the WST can be reliably used as a powerful estimator in modern analyses of LSS data \footnote{We note that this result relies on the adoption of a reliable and accurate HOD model.}. 

We should take note, at this point, of certain limitations of our current analysis, that we plan to tackle in a follow-up work. First, the Taylor expansion from Eq. \eqref{Xtaylor}, that we used to emulate the parameter dependence of our model vector in the likelihood (Eq. \ref{eq:LogL}) is inevitably expected to break down in parameter regions far away from the fiducial cosmology, a fact that essentially prevents us from capturing any substantial non-Gaussianities present in the likelihood. In order to do this, a full model for the non-linear dependence of the WST coefficients as a function of the cosmological parameters will need to be developed, similar to the training procedure of an emulator for a given clustering statistic. We envision that well-established emulation techniques (such as the one presented in Ref. \citep{Yuan:2022jqf} for the correlation function, or emulators at the field level \citep{Kaushal:2021hqv}) could be straightforwardly expanded to enable a full WST application, such as the one we performed in this work. Second, the Hubble parameter was not explicitly varied in our analysis, but rather obtained as a derived parameter through the fixed angular scale $\theta_{\star}$ (to the value measured by the {\it Planck} satellite \citep{Planck2018}) in the ${\tt ABACUSSUMMIT}$ simulations. This limitation can be easily overcome by using a different set of mocks, in which $h$ is explicitly varied. Furthermore, the current set of {\tt ABACUS} mocks we used did not account for the effect of light-cone evolution of the galaxy clustering within the survey volume\footnote{This effect is already captured, however, by the covariance from the {\tt PATCHY} mocks.}, an effect that can be rather easily overcome using the next generation of {\tt ABACUS} mocks already in production. Lastly, the weighting scheme \eqref{eq:wc} and \eqref{eq:fkpweight}, which we adopted to correct for the data systematics, was designed for the power spectrum case, rather than for the WST. Even though this option is expected to capture these effects to some extent, a correction scheme tailored to the WST estimator would be preferred. After all of the above improvements are implemented, a fair comparison against the results obtained by state-of-the-art re-analyses of BOSS data using perturbation theory (\citep{Ivanov_2020,d_Amico_2020,Philcox_2020,PhysRevD.105.043517,Chen_2022,Zhang_2022,2022arXiv220410392C}) or emulators (\citep{Nishimichi_2019,Miyatake:2020uhg,PhysRevD.102.063504,2021arXiv211006969K,Zhai:2022yyk,Yuan:2022jqf}) will finally be possible, a step that we reserve for future work. 

Our BOSS analysis hints at a wide range of exciting future applications of the WST in the context of LSS cosmology. In our previous work \citep{Valogiannis:2021chp}, we showed that the WST coefficients are particularly sensitive to the properties of massive neutrinos, thanks to their innate ability to capture clustering information beyond the traditional 2-point function, combined with their ability to trace the properties of voids. Even though we did not include neutrinos in our current analysis, since they were not varied in the {\tt ABACUSSUMMIT} simulations, they can be easily incorporated using a future set of mocks that captures their effect, which could then potentially allow us to obtain powerful constraints on the sum of the neutrino masses. This is also the case for a variety of other $\Lambda$CDM extensions, such as theories for modified gravity or dynamical dark energy.
We also emphasize here that the procedure we laid out is very flexible, and can be straightforwardly applied to any future set of spectroscopic galaxy observations, given an associated set of systematic weights and mock catalogues, both of which are commonly produced to support analyses using traditional estimators. For example, and subject to the additional improvements discussed above, our framework can be easily adjusted for a future application to spectroscopic observations by DESI. 

We note that despite their impressive performance in the context of an LSS analysis, the wavelets we used to implement the scattering network from Eq. \eqref{eq:WSTcoeff:sol} were proposed in the context of a 3D molecular chemistry application \citep{10.5555/3295222.3295400,doi:10.1063/1.5023798}. One could envision developing wavelets optimized for a cosmological application, which can further improve the benefits of a WST analysis. For example, equivariant wavelets \citep{2021arXiv210411244S} can find a natural application in the case of fields with a particular directional dependence, as is the galaxy overdensity observed in redshift space, that we have used as the input field in the current work.

In addition, we comment on the fact that the second order WST used in this analysis encodes information from correlation functions up to 4th order.  It would be interesting, as a result, to compare a BOSS analysis using WST and the 4-point correlation function \citep{Philcox:2021hbm} in future work.

Through this first application of the wavelet scattering transform on actual galaxy observations, we demonstrate that this technique can serve as a promising tool for current and future applications of cosmological parameter inference. 

%%%%%%%%%%%%%%%%%%%%%%%%%%%%%%%%%%%%%%%%%%%%%%
\vspace{0.5cm}
\subsection*{Acknowledgments} 
CD is partially supported by NSF grant AST-1813694.
This work is supported by the National Science Foundation under Cooperative Agreement PHY-2019786 (The NSF AI Institute for Artificial Intelligence and Fundamental Interactions).
We would like to thank Sihan Yuan for kindly sharing the first generation of {\tt ABACUSSUMMIT} mocks with us, for providing help with the use of {\tt ABACUSHOD} and for discussions over the course of this work. We are grateful to Boryana Hadzhiyska, Lehman H. Garrison and Daniel J. Eisenstein for valuable discussions and information on the suite of {\tt ABACUSSUMMIT} simulations and the associated galaxy mocks. We would also like to thank Arthur Tsang for detailed comments on our draft. Furthermore, we would like to thank Florian Beutler and Hector Gil Marin for useful discussions on the BOSS data and also Sihao Cheng and Tom Abel for sharing insights on the wavelet scattering transform. 

The massive production of all MultiDark-Patchy mocks for the BOSS Final Data Release has been performed at the BSC Marenostrum supercomputer, the Hydra cluster at the Instituto de Fısica Teorica UAM/CSIC, and NERSC at the Lawrence Berkeley National Laboratory. That work acknowledges support from the Spanish MICINNs Consolider-Ingenio 2010 Programme under grant MultiDark CSD2009-00064, MINECO Centro de Excelencia Severo Ochoa Programme under grant SEV- 2012-0249, and grant AYA2014-60641-C2-1-P. The MultiDark-Patchy mocks was an effort led from the IFT UAM-CSIC by F. Prada's group (C.-H. Chuang, S. Rodriguez-Torres and C. Scoccola) in collaboration with C. Zhao (Tsinghua U.), F.-S. Kitaura (AIP), A. Klypin (NMSU), G. Yepes (UAM), and the BOSS galaxy clustering working group.

\newpage

%%%%%%%%%%%%%%%%%%%%%%%%%%%%%%%%%%%%%%%%%%%%%%
\appendix

\section{WST for masked density fields}\label{sec:App_WSTmask}
In the original WST implementation \eqref{eq:WSTcoeff:power} in {\tt KYMATIO}, the input field is assumed to be a periodic 3D cube, such as, for example, the output of an N-body simulation. Given that in this application, however, we work with data that occupy the non-trivial survey geometry of BOSS, we need to make the necessary modifications. In particular, and in direct analogy to the power spectrum case, we start by embedding the masked density fields \eqref{eq:deltamasked} (both from the data and the cut-sky mocks) into 3D cubes using {\tt nbodykit}. These 3D grids, which contain both the actual volume of the survey and also the part of the cube that lies outside of the BOSS mask, are fed as input into {\tt KYMATIO}. We then need to make sure that the fundamental WST evaluations, wavelet convolution, modulus and averaging, include only the contributions from regions within the mask. In practice, we modify {\tt KYMATIO} such that regions of the 3D grid that lie outside the mask do not contribute to the convolutions in Eq. \eqref{eq:WSTcoeff:power}, in any order. This is straightforward to implement with minimal modifications, given the exact knowledge of the survey binary mask. Likewise, the regions outside the mask are always zeroed out and do not receive any contributions from the density field through the wavelet convolutions. Finally, we take the modulus and average over the parts of the field that lie within the survey footprint, in order to get the WST coefficients from a masked input field. This procedure is analogous to the corresponding evaluation of the power spectrum monopole for an input masked field. 

In addition to occupying an irregular survey geometry, we also note that the density fields we work with in this application are anisotropic, due to the effects of RSD, with the survey line-of-sight lying along the radial direction in a spherical coordinate system. Given that the basis of solid harmonic wavelets we adopted was designed for an isotropic input field, without treating any direction as special, the current WST configuration might not fully leverage all the information encoded in the RSD field (similar to evaluating only the monopole of the power spectrum, that averages over all directions). Such a shortcoming is indeed possible to overcome, for example with the directional-dependent equivariant wavelets of Ref. \citep{2021arXiv210411244S}. 
We defer this study for future work.

\section{Numerical Convergence}\label{sec:App_convergence}

\begin{figure}[ht!]
\includegraphics[width=0.49\textwidth]{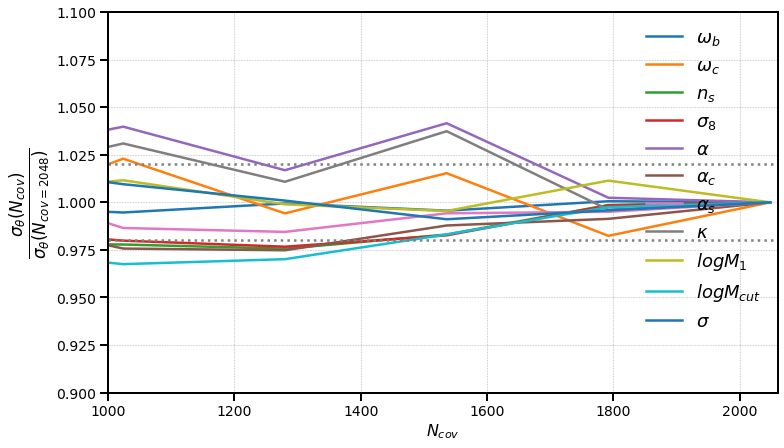}
\caption{The 1-$\sigma$ errors on the cosmological parameters, $\sigma_{\theta}$, plotted as a function of the number of {\tt PATCHY} mock realizations, $N_{cov}$, used to evaluate the WST covariance matrix. The y axis is normalized with respect to the 1-$\sigma$ errors obtained when using the total number of available $N_{cov}=2048$ realizations.}
\label{fig:WST_convergence_cov} 
\end{figure}

Given that the covariance matrix of the WST coefficients is evaluated from simulations, we need to make sure that the number of realizations used is sufficient to guarantee the numerical convergence of the results. We show in Fig. \ref{fig:WST_convergence_cov} the $1-\sigma$ errors on the cosmological parameters as a function of the number of {\tt PATCHY} mock realizations used to evaluate the covariance matrix. We find that the change (relative to the results obtained from the full suite of $N_{cov}=2048$ realizations) is smaller than $1.8 \%$ for all parameters, when using $N_{cov}\geq 1800$ realizations. This confirms the numerical convergence of the WST covariance.

\section{Data Compression}\label{sec:App_compression}

In addition to the covariance matrix convergence discussed in the previous section, we also confirm the numerical stability of the derivatives entering the Taylor expansion in Eq. \eqref{Xtaylor}. After testing how the 1-$\sigma$ errors obtained on the cosmological parameters change as a function of the number of realizations used to evaluate the derivatives, we found variations of at most 6$\%$ when using half of the total realizations. Furthermore, we note again that the simulations used to construct the derivatives were run with phase-matched initial conditions. This fact, in combination also with the much higher volume and resolution of {\tt ABACUS} compared to other existing simulations, is expected to modulate the noise due to cosmic variance, at least to some extent. Given, however, the relatively low number of 20 HOD realizations available for the evaluation of these derivatives, we also test the numerical stability by repeating our analysis using a compressed data vector. 

In particular, for a likelihood $\mathcal{L}(\theta|\bold{d})$, the quantity \citep{2018MNRAS.476L..60A}
\begin{equation}\label{eq:compt}
t = \nabla_{\theta}\mathcal{L}(\theta|\bold{d}),
\end{equation} 
corresponds to a compression from the original data vector of dimensionality $N_{d}$ down to one with dimensions equal to the number of parameters, $n$. In the case of a Gaussian likelihood with a data covariance independent of the cosmological parameters, Eq. \eqref{eq:compt} is further simplified to 
\begin{equation}\label{eq:comptLin}
t = \nabla_{\theta}\bold{X}^T C^{-1} \left[\bold{X}_\bold{d}-\bold{X}_t(\theta_{\rm fid})\right],
\end{equation} 
which represents a linear and lossless compression that preserves the Fisher matrix of the original estimator \citep{Tegmark_1997,2000MNRAS.317..965H}. The Gaussian likelihood of the compressed statistic is 
\begin{equation}\label{eq:Logt}
\log \mathcal{L}_{t}(\theta|\bold{d}) = -\frac{1}{2} \left[t-t(\theta)\right]^{\rm T} C_t^{-1}\left[t-t(\theta)\right] + {\rm const.},
\end{equation} 
where 
\begin{equation}\label{eq:compttheta}
t(\theta) = \nabla_{\theta}\bold{X}^T C^{-1} \left[\bold{X}_t(\theta)-\bold{X}_t(\theta_{\rm fid})\right],
\end{equation} 
and with $C_t$ the covariance matrix of t. In the particular case of the WST, the original data vector of $N_{d}=76$ coefficients is compressed down to $n=11$ numbers. Reductions of this kind greatly reduce the challenges associated with parameter inference from high-dimensional data vectors and have been utilized in bispectrum applications \citep{2019MNRAS.484.3713G,2021PhRvD.103d3508P,2022MNRAS.tmp.2216B}. More importantly for our case, the compression \eqref{eq:compt} leads to a statistic that is less sensitive to numerical noise, being a weighted average of the original data points. Indeed, such a compression has been recently used to accelerate the convergence of Fisher forecasts, the numerical stability of which is notoriously challenging in the case of noisy derivatives \citep{Coulton:2022rir,Coulton:2022qbc,Jung:2022rtn}. 

\begin{figure}[ht!]
\includegraphics[width=0.49\textwidth]{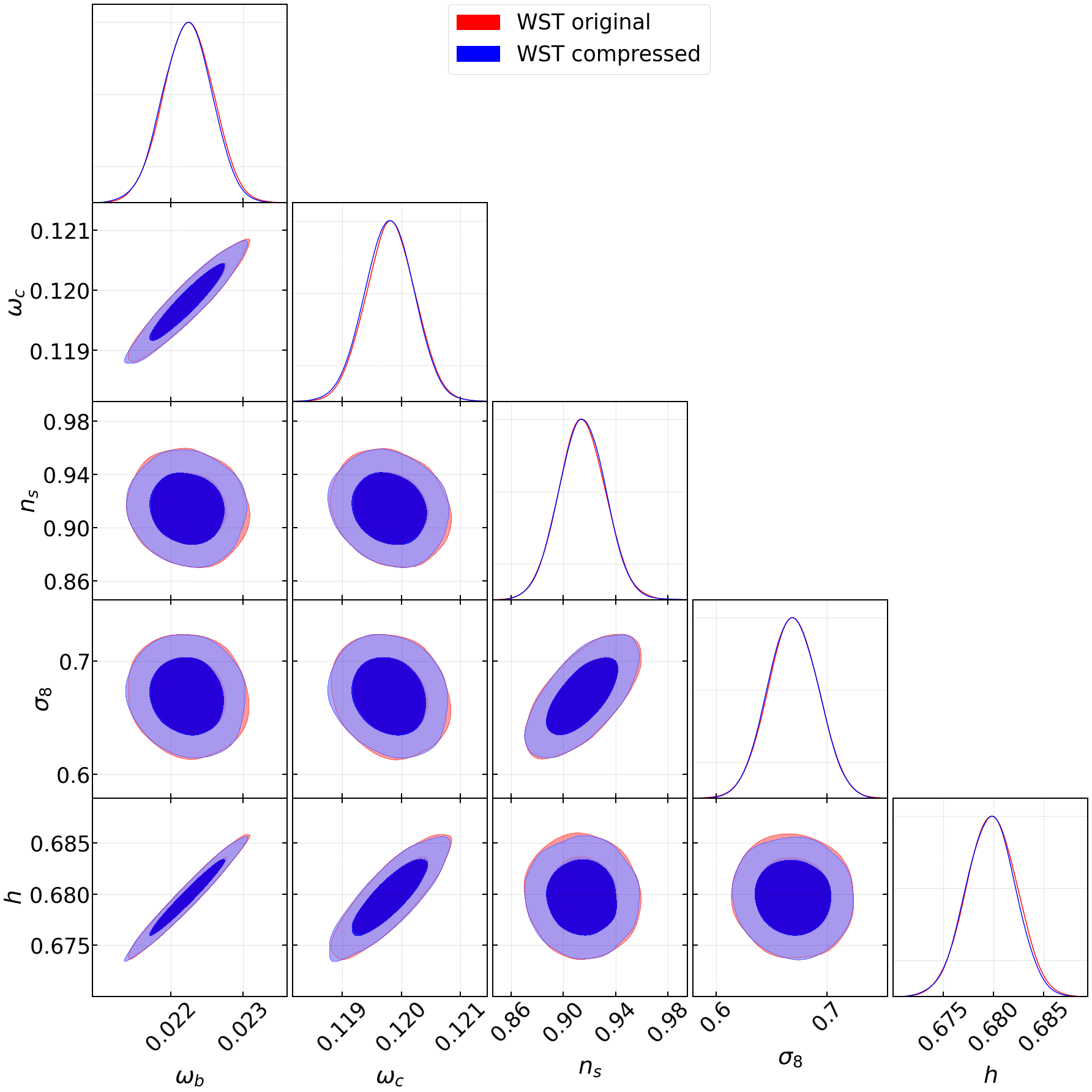}
\caption{Constraints on the cosmological parameters obtained from the original WST coefficients defined in \S\ref{sec:WSTeval} (red contours), as well as from the compressed version of the WST data vector from Eqs. \eqref{eq:comptLin}-\eqref{eq:compttheta} (blue contours). The results shown above were obtained after imposing a BBN Gaussian prior on the value of $\omega_b = 0.02268 \pm 0.00038$.}
\label{fig:WSTcomp}
\end{figure}

\begin{figure}[ht!]
\includegraphics[width=0.49\textwidth]{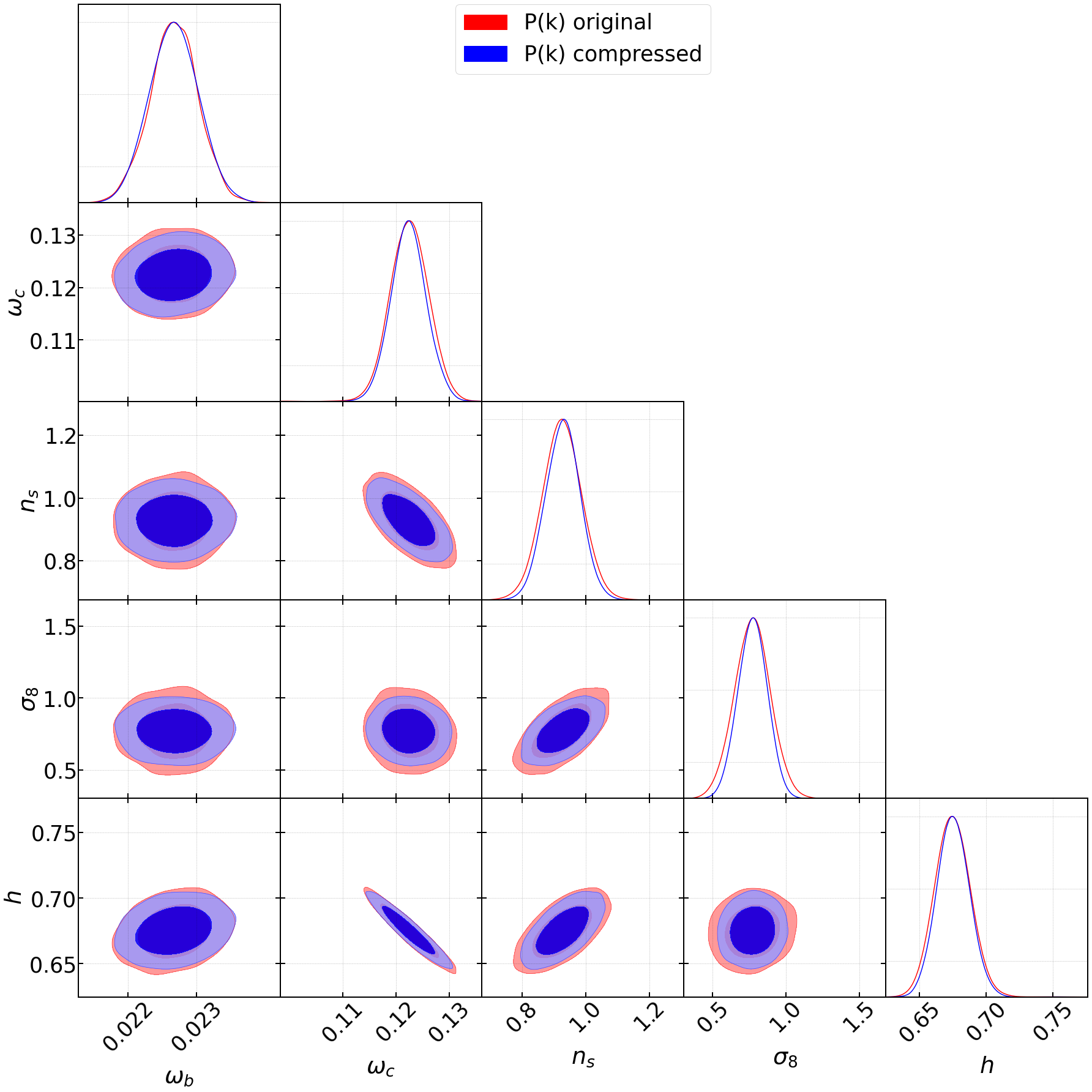}
\caption{Constraints on the cosmological parameters obtained from the galaxy power spectrum multipoles (red contours), as well as from the compressed version of the power spectrum data vector from Eqs. \eqref{eq:comptLin}-\eqref{eq:compttheta} (blue contours). The results shown above were obtained after imposing a BBN Gaussian prior on the value of $\omega_b = 0.02268 \pm 0.00038$.}
\label{fig:Pkcomp}
\end{figure}

We compress the WST and the power spectrum multipole vectors using Eqs. \eqref{eq:comptLin}-\eqref{eq:compttheta} and then repeat the parameter inference application of the analysis section, sampling from the likelihood \eqref{eq:Logt}. In Fig. \ref{fig:WSTcomp}, we compare the marginalized 2-dimensional posteriors obtained for the cosmological parameters using the compressed WST, against the results obtained from the original, un-compressed, WST analysis in the case of a BBN prior. The very small differences between these two sets of results serve as additional confirmation of the robustness of the main analysis. Finally, in Fig. \ref{fig:Pkcomp} we show the same comparison for the multipoles of the galaxy power spectrum, reaching a similar conclusion.

\section{Full Parameter Space for P(k) and WST}\label{sec:App}

\begin{figure*}[ht!]
\includegraphics[width=0.99\textwidth]{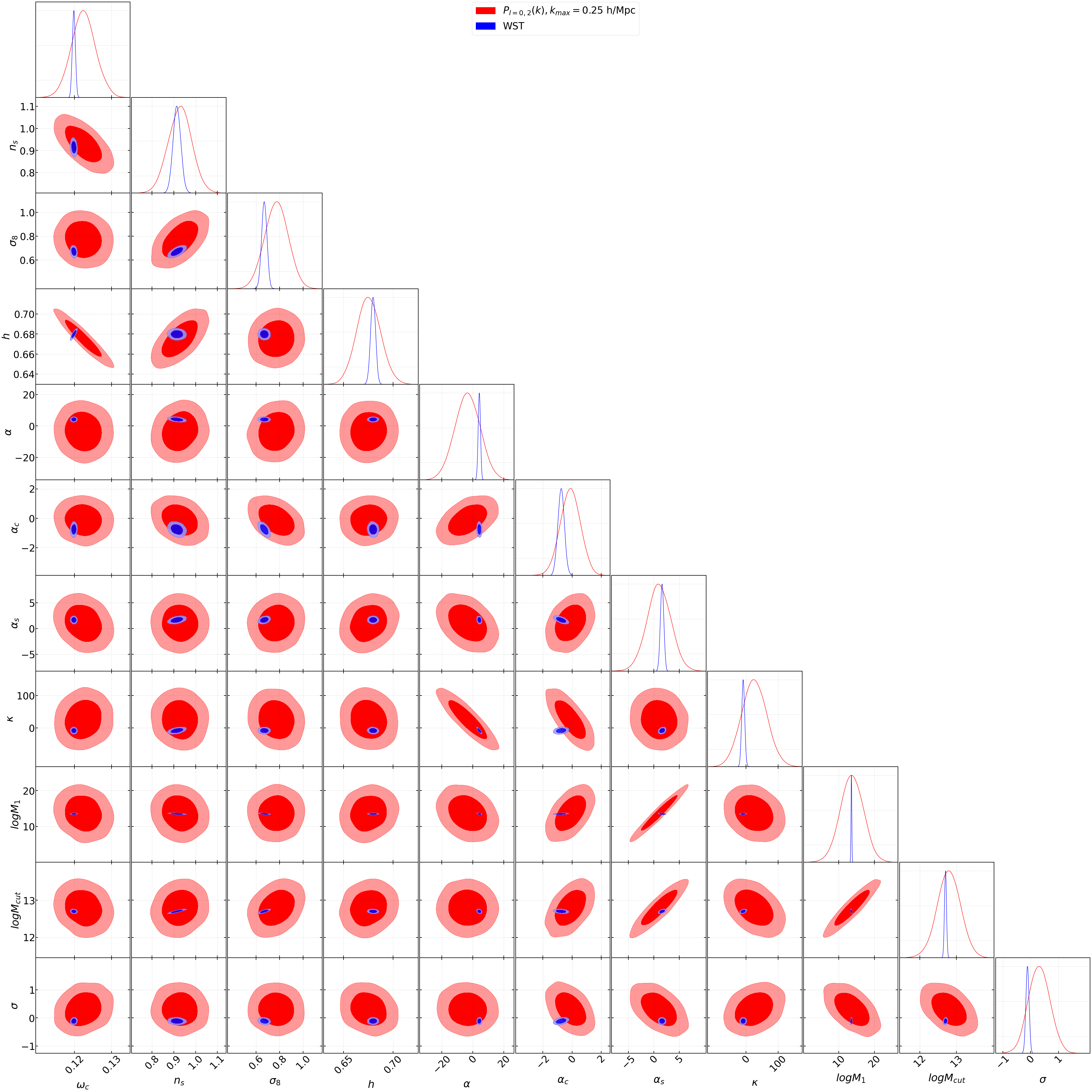}
\caption{Same as in Fig.~\ref{fig:BBNpriorOmegab_WST_vs_Pk}, but now including the 7 parameters of the HOD model. $M_{\rm cut}$ and $M_1$ are expressed in units of $M_{\odot}/h$.}
\label{fig:BBNpriorOmegab_WST_vs_PkALLHOD} 
\end{figure*}

\begin{figure*}[ht!]
\includegraphics[width=0.99\textwidth]{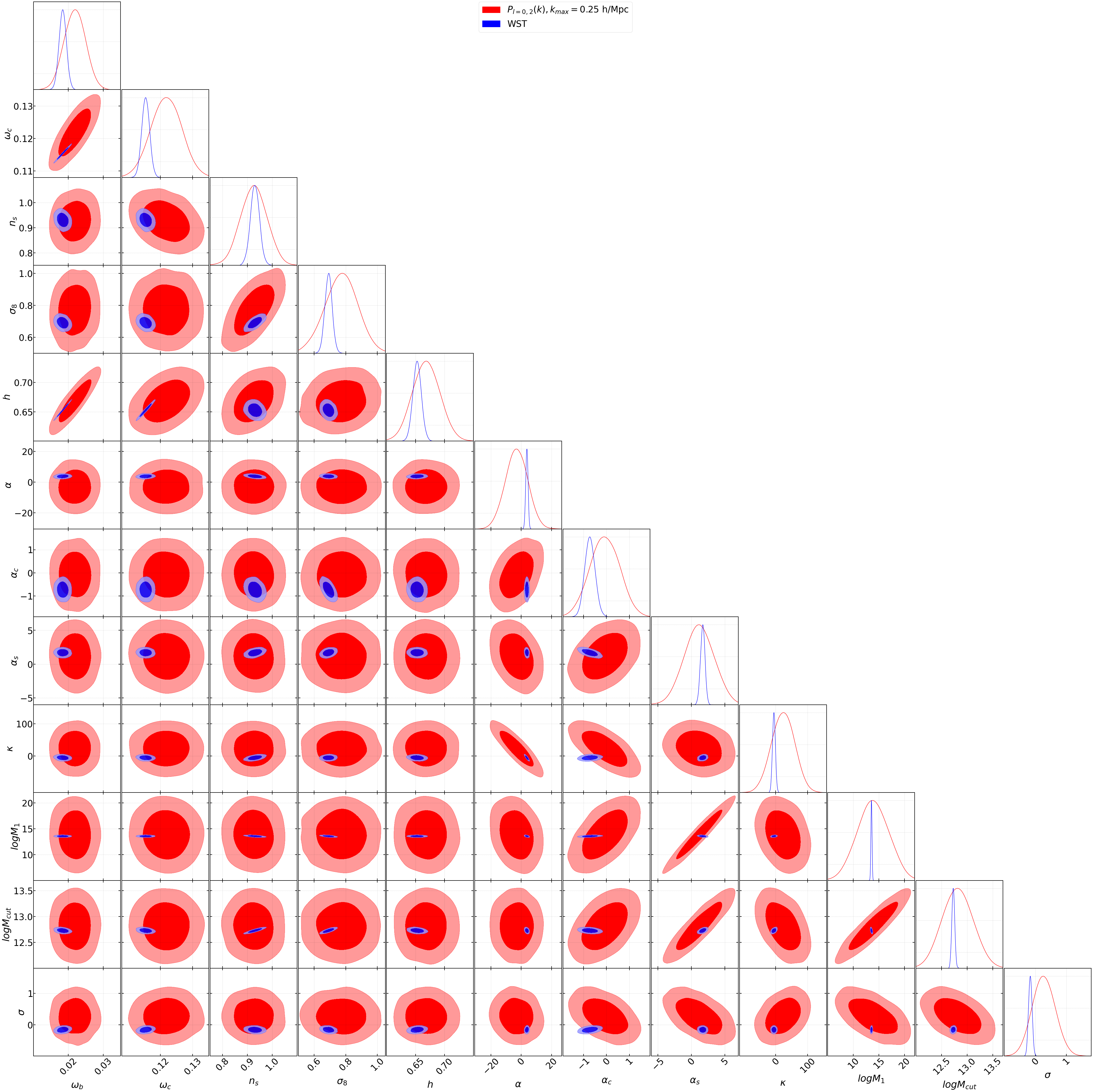}
\caption{Same as in Fig.~\ref{fig:flatpriorOmegab_WST_vs_Pk}, but now including the 7 parameters of the HOD model. $M_{\rm cut}$ and $M_1$ are expressed in units of $M_{\odot}/h$.}
\label{fig:flatpriorOmegab_WST_vs_PkALLHOD} 
\end{figure*}

For completeness, in this appendix we show the full corner plots from our likelihood analysis, including the marginalized posteriors of the 7 parameters of our HOD model, that we treated as nuisance parameters. In particular, in Fig.~\ref{fig:BBNpriorOmegab_WST_vs_PkALLHOD} we show the full corner plot of the analysis with a BBN prior on the value of $\omega_b$, while Fig.~\ref{fig:flatpriorOmegab_WST_vs_PkALLHOD} illustrates the results for the case of unrestricted priors on all parameters. We see that the 12 parameters shown are consistent (within $1\sigma$) for the mean values recovered using the WST coefficients and the power spectrum multipoles.

\newpage

\bibliographystyle{apsrev4-2_16.bst}

%\bibliography{main.bib}

\end{document}